\PassOptionsToPackage{sort&compress,numbers}{natbib}
\documentclass[numbers]{els-mrw} 

\usepackage{amsmath,amssymb,amsfonts,amsthm,makeidx,graphicx}
\usepackage{txfonts}
\usepackage{helvet}
\usepackage{graphicx}
\usepackage{hyperref}
\usepackage{color}
\usepackage[dvipsnames]{xcolor}
\graphicspath{{./figures/}}

\hypersetup{
    pdfnewwindow=true,      
    colorlinks=true,       
    linkcolor=blue,          
    citecolor=blue,        
    filecolor=blue,      
    urlcolor=blue        
}

\begin{document}

\chapter{Neutrino Physics and Astrophysics at Colliders}\label{chap1}

\author[1,2]{Bei Zhou\footnotemark[2]}
\author[1]{Pedro Machado\footnotemark[1]}
\address[1]{\orgdiv{Theory Division}, \orgname{Fermi National Accelerator Laboratory}, \orgaddress{Batavia, Illinois 60510, USA}}
\address[2]{\orgdiv{Kavli Institute for Cosmological Physics}, \orgname{University of Chicago}, \orgaddress{Chicago, Illinois 60637, USA}}

\articletag{Chapter Article tagline: update of previous edition, reprint.}

\maketitle


\begin{abstract}[Abstract]
Nonzero neutrino masses guarantee new physics and neutrinos are excellent probes of extreme environments in the Universe. 
The recent collider neutrino experimental program, including FASER$\nu$ and SND@LHC, along with the planned Forward Physics Facility at the High-Luminosity Large Hadron Collider, is opening a new window into neutrino physics and astrophysics. 
In this article, we review recent achievements and prospects of collider neutrino experiments, including key achievements such as the first measurements of collider neutrino interactions at unprecedented energies and the exploration of new physics scenarios, like dark matter candidates, sterile neutrinos, and non-standard neutrino interactions. 
For concreteness, we will focus on the significant scientific opportunities presented by the Forward Physics Facility, which will enable precision measurements of neutrino cross sections and proton structure at low parton momentum fraction. 
Furthermore, collider neutrino studies will substantially reduce systematic uncertainties in calculating atmospheric neutrino fluxes, thereby improving astrophysical neutrino observations as well as advancing our understanding of cosmic-ray interactions.
\end{abstract}

\begin{keywords}
 	Neutrino Physics\sep 
	Neutrino Astrophysics\sep 
	Neutrino Interactions \sep 
	Physics Beyond the Standard Model \sep
	Collider 
\end{keywords}

\vfill
\footnotetext[1]{\href{mailto:beizhou@fnal.gov}{beizhou@fnal.gov}}
\footnotetext[2]{\href{mailto:pmachado@fnal.gov}{pmachado@fnal.gov}}

\begin{glossary}[Nomenclature]
	\begin{tabular}{@{}lp{34pc}@{}}
		ALP & Axion-Like Particle\\
		ATLAS & A Toroidal LHC ApparatuS\\
		BSM & Beyond the Standard Model\\
		CC & Charged Current\\
        CCDIS & Charged-Current Deep-Inelastic Scattering \\
		DIS & Deep-Inelastic Scattering\\
		EAS & Extensive Air Showers\\
		ECC & Emulsion Cloud Chamber\\
		FASER & ForwArd Search ExpeRiment\\
		FASER$\nu$ & FASER-neutrino\\
		FLArE & Forward Liquid-Argon Experiment\\
		FPF & Forward Physics Facility\\
		HL-LHC & High-Luminosity LHC\\
		HNL & Heavy Neutral Lepton\\
        LArTPC & Liquid Argon Time Projection Chamber \\
		LHC & Large Hadron Collider\\
		NC & Neutral Current\\
        NCDIS & Neutral-Current Deep-Inelastic Scattering \\
		NSI & Non-Standard Interaction\\
		PDF & Parton-Distribution Function\\
		QES & Quasi-Elastic Scattering\\
		RES & REsonant Scattering  \\
		SND@LHC & Scattering and Neutrino Detector at the LHC\\
		TPC & Time Projection Chamber\\
		WBP & neutrino-nucleus W-Boson Production\\
	\end{tabular}
\end{glossary}

\section*{Objectives}
\begin{itemize}
	\item Understand the motivation and significance of neutrino physics and astrophysics at colliders
	\item Learn about experimental facilities that pioneered collider neutrino detection and the planned Forward Physics Facility
	\item Explore neutrino production mechanisms in colliders and the resulting neutrino flux characteristics
	\item Examine the physics of neutrino interactions at TeV-scale energies and rare processes
	\item Learn about the potential for discovering Beyond-the-Standard-Model physics using collider neutrinos
	\item Connect collider neutrino measurements with observations of atmospheric neutrinos, astrophysical neutrinos, and cosmic rays
\end{itemize}

\section{Introduction}\label{intro}

Neutrinos, despite being among the most abundant particles in the universe, remain challenging to study due to their exceptionally small interaction probabilities with matter. However, studying neutrinos is extremely important. 
Neutrinos guarantee new physics as their masses cannot be explained by the Standard Model, so we must scrutinize all the possibilities that can be measured in the neutrino sector to look for BSM physics.  
Astrophysical neutrinos are excellent probes of dense and other extreme environments in the Universe; they are the smoking gun of the origin of cosmic rays --- a century-long problem; and they are also powerful probes of new physics.

When protons collide at the LHC, a significant flux of high-energy neutrinos is produced in the very forward direction ($\theta < 1$ mrad sideways).
By placing detectors far from the interaction point, we can study these neutrinos while avoiding the high-multiplicity backgrounds from proton-proton collisions.
This program began with FASER~\cite{FASER:2019dxq}, a compact detector designed to search for light, long-lived particles, and can also detect neutrinos, in the forward region of the LHC during Run 3. 
FASER was complemented by FASER$\nu$~\cite{FASER:2020gpr}, a dedicated subdetector for neutrino detection.
In particular, FASER achieved the first observation of collider neutrinos~\cite{FASER:2023zcr}.
The success of both experiments demonstrated rich physics potentials in the far-forward region and motivated the larger FPF program~\cite{Feng:2022inv, FPF_web}. 
FPF will capitalize on this opportunity by hosting a suite of experiments approximately 620m downstream of the ATLAS interaction point. 
It will address several fundamental questions in particle physics and astrophysics that have remained elusive with conventional experiments, thanks to its enhanced detection capabilities, larger detectors, and improved angular coverage, which maximize the physics potential of forward-going particles at the LHC.

First, neutrino cross sections at TeV energies are largely unmeasured, with the highest energy direct measurements coming from fixed-target experiments at $\sim$350 GeV~\cite{CCFR:1997tam, NuTeV:2003kth, NOMAD:2009qmu}. 
While the Standard Model predicts how these cross sections should scale with energy, new physics effects could modify this behavior at higher energies~\cite{Jain:2000pu, Arguelles:2015wba, Becirevic:2018uab, Bai:2025pef}. 
The FPF will measure neutrino interactions from $\sim$100 GeV to several TeV, bridging the gap between measurements from fixed-target experiments and those from atmospheric and astrophysical neutrinos~\cite{Ariga:2025qup}.

Second, the FPF will probe the structure of the proton~\cite{Amoroso:2022eow, Cruz-Martinez:2023sdv} in a previously inaccessible region. 
Through neutrino production and interactions, we can study PDFs down to very small values of the momentum fraction $x$ and up to high momentum transfer $Q^2$. 
This kinematic region is particularly challenging to access with other experiments, yet it is crucial for understanding the physics at the energy frontier and for interpreting cosmic-ray observations.

Third, FPF's unique configuration makes it sensitive to various potential new-physics signatures. 
The facility could detect light dark matter particles produced in the forward region, identify new forces that primarily affect neutrinos, or discover other weakly-interacting particles that might have escaped detection in central detectors.

Last but not least, the facility will significantly advance astrophysics and astroparticle physics. 
The FPF will significantly reduce the major uncertainties in neutrino telescope measurements. 
TeV--PeV atmospheric neutrinos, produced by cosmic-ray interactions in the atmosphere, constitute the main background for astrophysical neutrino measurements. 
These neutrinos originate primarily from the decay of mesons (pions, kaons, and charm hadrons), but their production cross sections in the relevant kinematic region are poorly constrained. 
The FPF will measure forward productions of these mesons directly, significantly improving predictions for atmospheric neutrino fluxes.
This measurement will be useful for reducing the uncertainties on both conventional and prompt atmospheric neutrino fluxes.
The latter is currently one of the largest sources of systematic uncertainties at neutrino observatories like IceCube, KM3NeT, and Baikal-GVD. 
Moreover, FPF will help with solving the ``muon puzzle'' problem in cosmic-ray measurements, a deficit of muons in cosmic-ray simulations with respect to experimental observations (see, e.g., Ref.~\cite{Albrecht:2021cxw}).

\section{Collider Neutrino Experiments}

\subsection{FASER and FASER$\nu$}

\begin{figure}[h!]
	\centering
	\includegraphics[width=.7\textwidth]{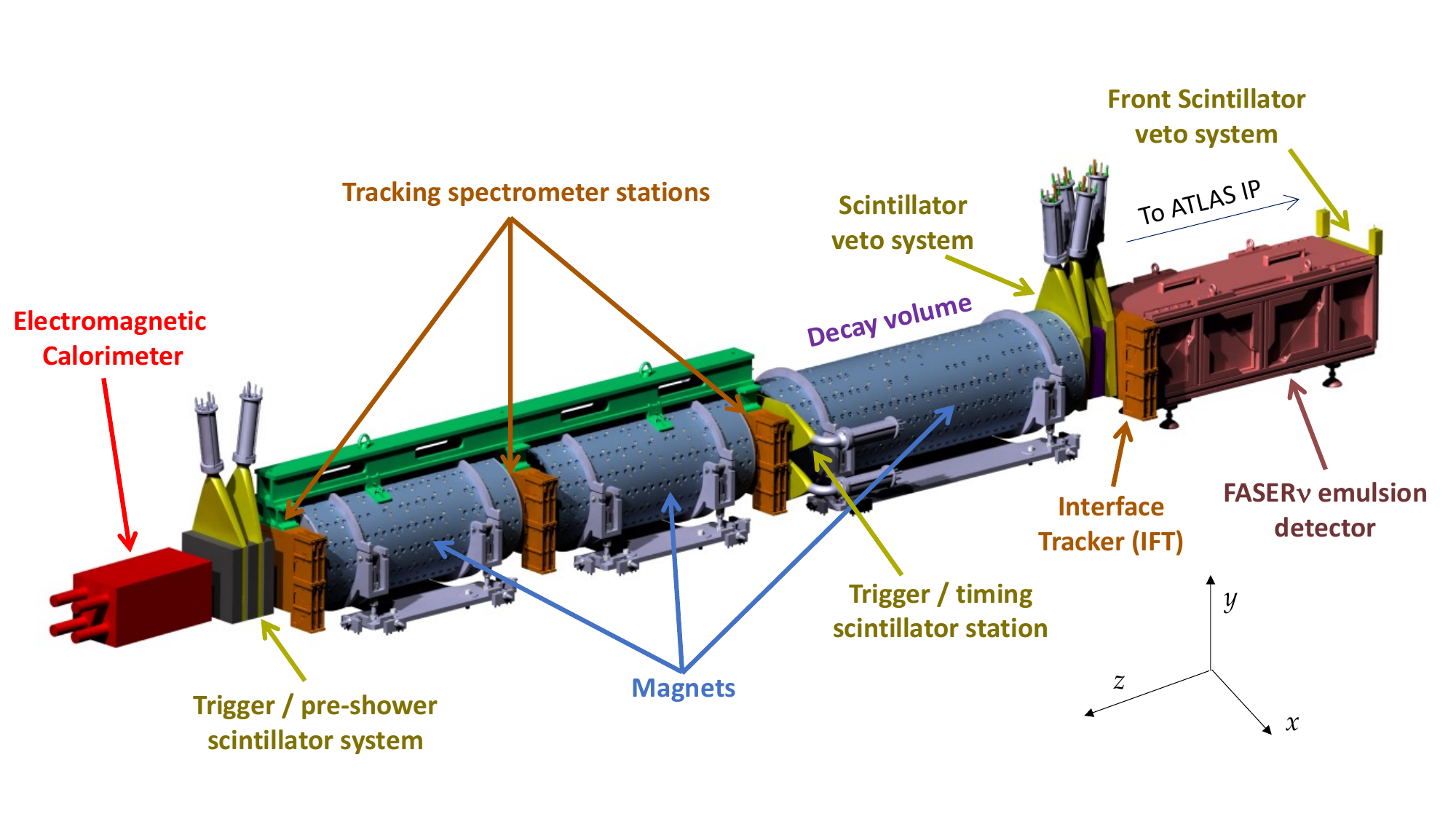}
	\caption{Layout of the FASER$\nu$ and FASER detectors, from Ref.~\cite{FASER:2022hcn}.}
	\label{Fig_FASERnu_experiment}
\end{figure}

Fig.~\ref{Fig_FASERnu_experiment} shows the layout of the FASER and FASER$\nu$ detectors. 
FASER$\nu$ is located directly in front of
FASER, 480 m from the ATLAS interaction point along the beam collision axis (to maximize the neutrino flux) in the tunnel TI12. 
FASER/FASER$\nu$ is the first experiment to detect neutrinos from a high-energy particle collider, and it aims to observe $\mathcal{O}(10^4)$ neutrino interactions around TeV energies and study all three neutrino flavors. 
The detector is composed of 770 emulsion layers interleaved with 1.1-mm-thick tungsten plates, with a total volume of 25~cm $\times$ 30~cm $\times$ 1.05~m and a total tungsten target mass of 1.1 tonnes. 
This detector design serves as both a target and a tracker: neutrinos interact in the dense tungsten, and the nuclear emulsion layers accurately record the resulting charged particle tracks with sub-micron precision. 
The emulsion films have a spatial resolution of approximately 300 nm~\cite{FASER:2025qaf}, allowing for the identification of short-lived tau lepton decays for $\nu_\tau$ detection. 
Because emulsions have no timing, the detector is periodically replaced (every $\sim$20--50~fb$^{-1}$) to avoid excessive track overlaps.

With 300 fb$^{-1}$ luminosity in Run 3, FASER$\nu$ is expected to record $\mathcal{O}(10^4)$ neutrino interactions.  
These include predominantly $\nu_\mu$, $\mathcal{O}(10^3)$ $\nu_e$, and a handful of $\nu_\tau$---the first tau neutrinos to be detected in a collider. 
By identifying charged leptons and event topologies, FASER$\nu$ can separately measure $\nu_e$/$\bar{\nu}_e$, $\nu_\mu$/$\bar{\nu}_\mu$, and $\nu_\tau$/$\bar{\nu}_\tau$ CC interaction cross sections up to TeV energies. 
Going forward, FASER$\nu$ will search for $\nu_\tau$ events by detecting tau lepton decays within the emulsion stack; even a few candidate $\nu_\tau$ events would be significant as no collider $\nu_\tau$/$\bar{\nu}_\tau$ has ever been observed. 
In addition, FASER can also detect neutrinos using the electronic detector~\cite{Arakawa:2022rmp}.
For more details of FASER$\nu$, see, e.g., 
Ref.~\cite{Feng:2017uoz} (original proposal of FASER), 
Ref.~\cite{FASER:2018bac} (FASER technical proposal), 
Ref.~\cite{FASER:2018ceo} (FASER letter of intent), 
Ref.~\cite{FASER:2019dxq} (FASER$\nu$ letter of intent), 
Ref.~\cite{FASER:2020gpr} (FASER$\nu$ technical proposal), 
Ref.~\cite{FASER:2022hcn} (FASER detector),
and Ref.~\cite{FASER:2025qaf} (FASER$\nu$ performance).

\subsection{SND@LHC (Scattering and Neutrino Detector at the LHC)}

SND@LHC is located 480 m downstream of the ATLAS interaction point, in the TI18 tunnel.
SND@LHC and FASER$\nu$ are situated to cover slightly different pseudorapidity regions, which enables complementary physics studies~\cite{SNDLHC:2022ihg}. 
FASER$\nu$ is on-axis, covering higher pseudorapidities ($\eta > 8.4$)~\cite{Arakawa:2022rmp}, and therefore is exposed to a larger neutrino flux and higher neutrino energies. 
SND@LHC is situated slightly off-axis, corresponding to a lower pseudorapidity range ($7.2<\eta<8.4$), which means a higher fraction of the neutrinos observed by the experiment originate from charm-hadron decay.

\begin{figure}[h!]
	\centering
	\includegraphics[width=.6\textwidth]{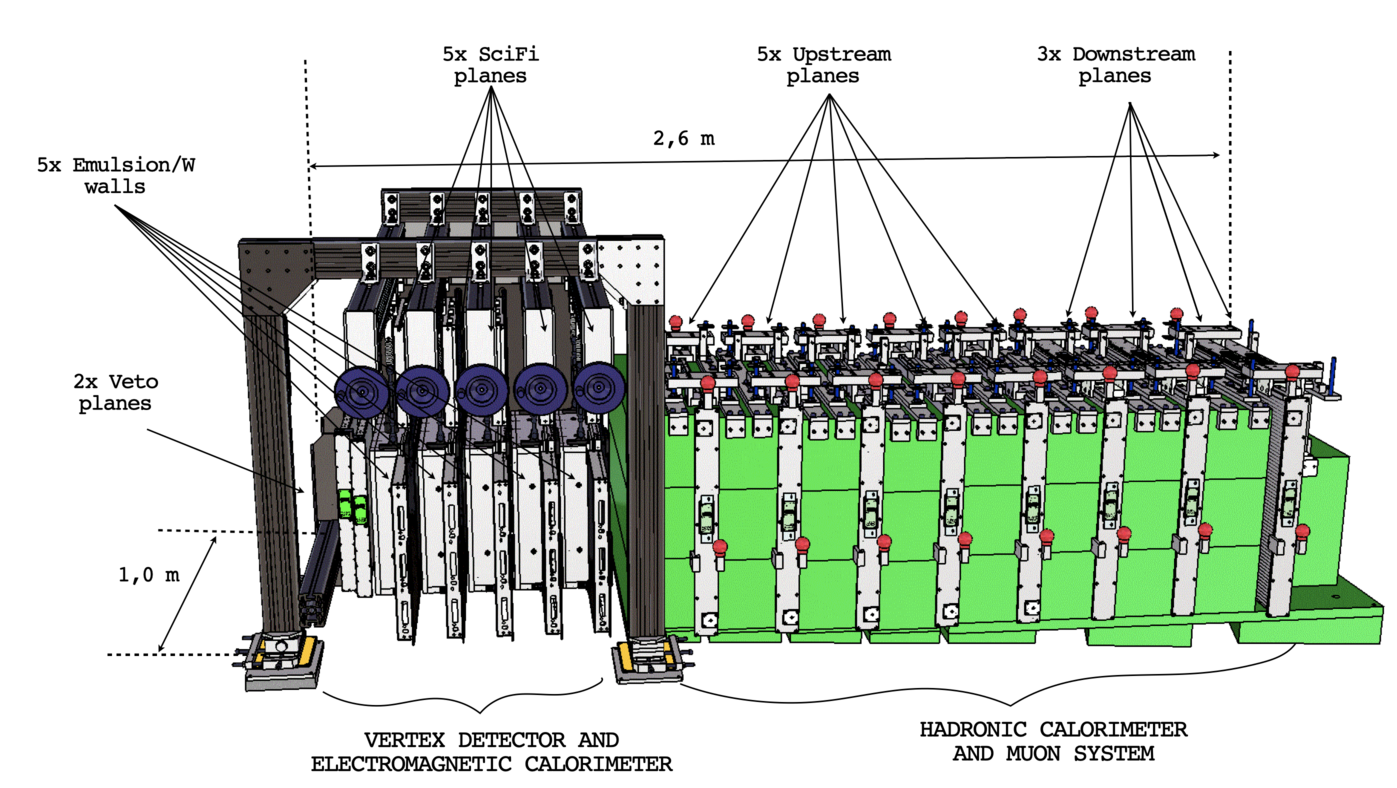}
	\caption{Schematic layout of the SND@LHC detector, from Ref.~\cite{SNDLHC:2022ihg}}
	\label{Fig_SND_layout}
\end{figure}

Figure~\ref{Fig_SND_layout} illustrates the layout of the SND@LHC detector. 
The detector consists of a hybrid system based on an 830-kg tungsten plate target interleaved with both emulsion and electronic trackers, which also serve as an electromagnetic calorimeter, and is followed by a hadronic calorimeter and a muon identification system.
The vertex detector consists of a series of Emulsion Cloud Chamber (ECC) walls: tungsten plates interleaved with nuclear emulsion films, providing a high-density target with micrometric tracking for interaction vertices. The target is segmented into five modular walls (totaling approximately 830 kg of tungsten) with emulsion films sandwiched between the plates. 
Between these walls are fast scintillating fiber tracking planes (SciFi) that timestamp the interactions and also act as an electromagnetic calorimeter for shower energy measurement. Upstream of the target, two veto scintillator planes tag incoming charged particles (primarily beam-induced muons) to reduce background. 
Downstream of the emulsion section, SND@LHC has a hadronic calorimeter and muon identification system, consisting of eight 20 cm-thick iron plates (green), each followed by one or two planes of 1 cm-thick scintillating bars.
The muon identification mainly consists of the last three scintillator bars. 
This configuration --- emulsion+SciFi target, followed by hadronic calorimeter and muon identification --- allows reconstruction of neutrino events and identification of the final-state lepton: e.g. a muon track in the muon system indicates a $\nu_\mu$ interaction, an electromagnetic shower in the emulsion+SciFi with no muon indicates a $\nu_e$ interaction, and a characteristically short tau decay topology would signal a $\nu_\tau$ interaction.

SND@LHC is expected to detect $\simeq 2000$ neutrino interactions during LHC Run 3. 
Simulation studies predict neutrino energies primarily in the $\sim100$~GeV--1~TeV range, originating largely from the decay of charm mesons. 
The experiment's design enables lepton flavor identification with $\sim 90\%$ efficiency for $\nu_e$ and $\nu_\mu$. 
With the excellent spatial resolution in the emulsions, it can detect charmed-hadron decays and identify tau decay kink signatures.
For more details on SND@LHC, see, e.g., Refs.~\cite{SNDLHC:2022ihg, SND_web}.

\subsection{Forward Physics Facility (FPF) at the High-luminosity LHC}

The FPF is a proposed underground cavern located approximately 620 m downstream of the ATLAS interaction point (IP) at the HL-LHC ($\sqrt{s}=14$ TeV, 3 ab$^{-1}$), designed to host far-forward experiments. 
With extensive shielding and exposure to the highest-energy forward particles, it will leverage a $\sim$20× increase in neutrino flux over Run 3 and a $\sim$20× increase in the target mass, enabling detection of millions of neutrino interactions, including thousands of $\nu_\tau$ events. 
Two complementary neutrino detectors are planned: the upgraded emulsion detector FASER$\nu$2 and the liquid-argon TPC FLArE~\footnote{The upgrade of SND@LHC was initially planned to be a part of FPF, named AdvSND~\cite{Feng:2022inv}, but is now planned instead as an independent detector at the current SND location, named SND@HL-LHC~\cite{Abbaneo:2926288}}. Together, they will deliver high-statistics measurements and search for new physics such as sub-GeV dark matter.
In the subsections below, we introduce each neutrino detector in the FPF. For more details, see, e.g., 
Ref.~\cite{Anchordoqui:2021ghd} (original proposal), 
Ref.~\cite{Feng:2022inv} (Snowmass white paper),
and Refs.~\cite{Adhikary:2024nlv, FPFWorkingGroups:2025rsc} (recent summaries and updates).
In addition to the officially planned FPF experiments, alternative mid-baseline proposals have also been put forward. For instance, Ref.~\cite{Kamp:2025phs} discusses the concepts of surface-based and underwater integrated neutrino experiments, which would utilize surface scintillator arrays and submerged water Cherenkov detectors in Lake Geneva to detect forward neutrinos from CMS and LHCb, respectively.

\subsubsection{FASER$\nu$2 (Emulsion Detector at FPF)}
FASER$\nu$2 is a scaled-up successor to the Run 3 FASER$\nu$ detector, located on-axis in front of the FASER2 spectrometer in the FPF. 
It adopts the same emulsion/tungsten ECC design, but with a $\simeq$20-ton target composed of $\simeq$3,300 emulsion films interleaved with two mm-thick tungsten plates, spanning $40\times40$ cm$^2$ in cross section and 6.6 m in length.

With 20× the mass and 20× the luminosity of FASER$\nu$, FASER$\nu$2 will collect $\mathcal{O}(10^6)$ neutrino interactions of all flavors. This enables precision cross-section measurements up to TeV energies. 
These data will enable a detailed study of physics both within the standard model (e.g., charm production and strange-quark PDF) and beyond (e.g., searching for light dark matter particles).

\subsubsection{FLArE (Forward Liquid-Argon Experiment)}
FLArE is a proposed LArTPC detector at the FPF of the HL-LHC. With a fiducial mass of about 10 tonnes, FLArE aims to record millions of neutrino interactions of all flavors. The design of LArTPC enables precise 3D particle tracking, identification, and calorimetry, with mm-scale spatial resolution and excellent electromagnetic shower containment. FLArE is particularly well-suited for detecting forward neutrinos and low-threshold dark matter interactions in the high-radiation HL-LHC environment. Its full active volume and high resolution make it capable of identifying event topologies and lepton flavor, especially for muons and electrons, while tau neutrino reconstruction remains more challenging.

To meet its physics goals, FLArE faces significant design and operational challenges, including the need for low-background triggering, high-rate data acquisition, and accurate event reconstruction in the presence of muon-induced noise. Key R\&D topics include the design of the time projection chamber (TPC), optimization of photon sensor systems for precise timing and triggering, and the development of machine learning-based trigger algorithms. Simulation studies indicate that FLArE will detect about 50 neutrino events per tonne per fb$^{-1}$, and machine learning tools such as convolutional neural networks and transformer models are proposed to enhance real-time data filtering and reduce trigger latency.


\section{Neutrino Fluxes from Colliders}

At high-energy colliders, neutrinos are predominantly produced through the decay of light hadrons generated at the interaction point. 
The high center-of-mass energies lead to copious meson and baryon production, particularly in the forward direction. 
After all prompt strong interaction decays take place, the remaining particles include pions, kaons, $D$ mesons, and various hyperons (baryons with valence strange quarks) like $\Lambda$, $\Sigma$, and $\Xi$. 
Pions contribute almost exclusively to the $\nu_\mu$ flux, since the branching ratio $\pi \to e\nu_e$ is suppressed by a factor of $(m_e/m_\mu)^2$. 
Charged kaons generate significant numbers of $\nu_\mu$ through $K \to \mu\nu_\mu$, and of $\nu_e$ via the semileptonic decay $K \to \pi e\nu_e$. 
$D$ mesons produce both $\nu_e$ and $\nu_\mu$ through channels like $D \to K e\nu_e$, while $D_s$ mesons serve as the primary source of tau neutrinos via $D_s \to \tau\nu_\tau$, followed by subsequent tau decays.

Forward hadron production can be estimated using various Monte Carlo event generators~\cite{Roesler:2000he, Ahn:2009wx, Ostapchenko:2010vb, Pierog:2013ria, Skands:2014pea, Sjostrand:2014zea, Fedynitch:2015kcn, Riehn:2015oba, Riehn:2017mfm, Fedynitch:2018cbl}. 
Since many of these hadrons have relatively long lifetimes, they may travel macroscopic distances before decaying. 
This necessitates accurate tracking of these particles throughout the experimental infrastructure~\cite{Battistoni:2015epi, Nevay:2018zhp, Kling:2021gos}. 
In principle, hadrons interacting with the collider components and tunnel walls could initiate hadronic showers, creating additional hadrons and thus a secondary neutrino flux. 
However, this secondary component contributes only a few percent to neutrinos above 100 GeV. For lower-energy neutrinos around 10 GeV or below, secondary production becomes significant but falls outside the energy range of interest for these facilities.

\begin{figure*}[htbp]
    \centering
    \includegraphics[width=\textwidth]{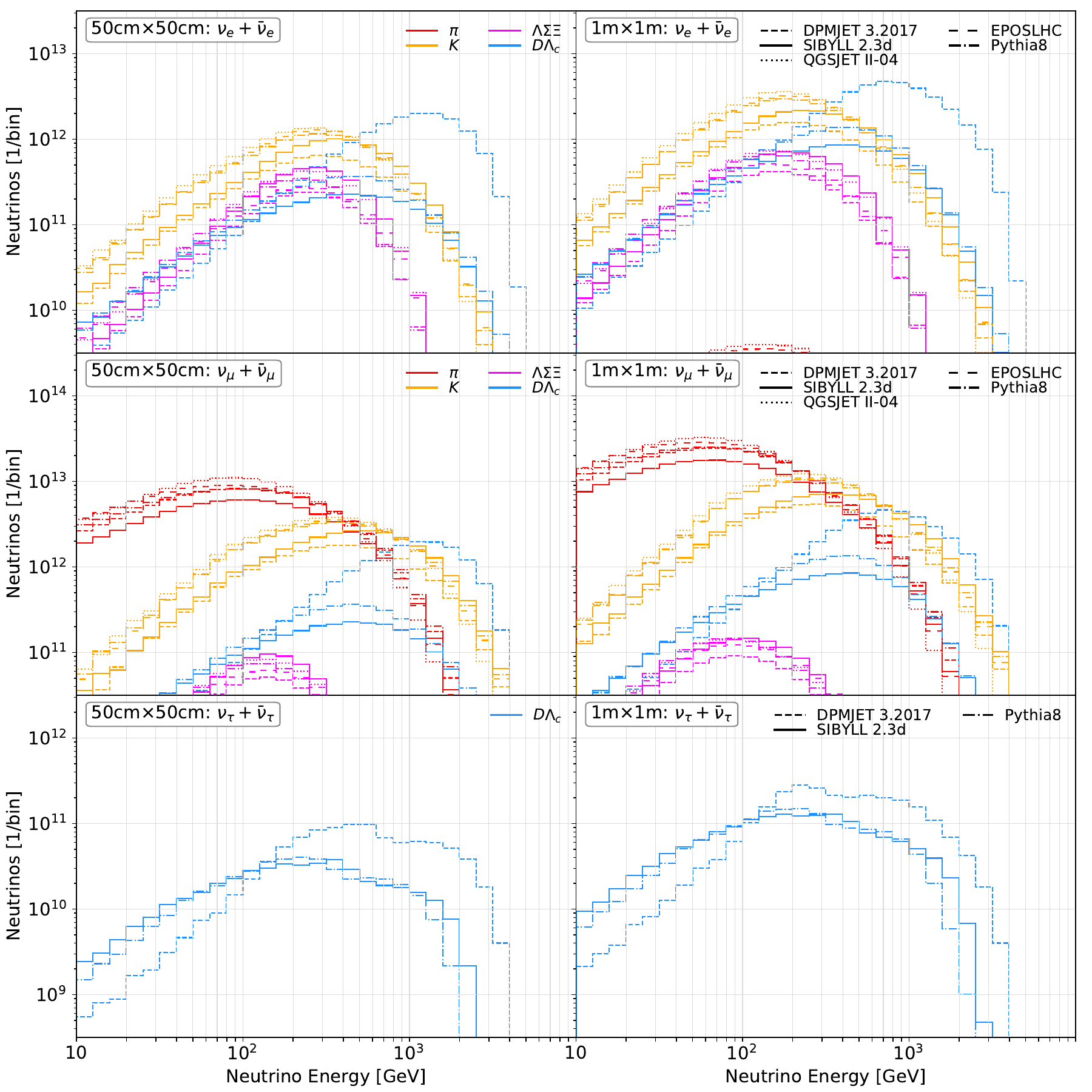}
    \caption{Predicted neutrino fluxes for each flavor (distributed vertically) at the FPF. 
    Parent hadrons are indicated by colors and the style of the histogram refers to various Monte Carlo generators. 
    Left and right panels indicate the neutrino fluxes for a cross-sectional area of 50~cm $\times$ 50~cm  and 1~m $\times$ 1~m at the FPF location at 620~m away from the beam interaction point.
    Figure taken from Ref.~\cite{Feng:2022inv}.}
    \label{fig:neutrino_flux}
\end{figure*}

Figure~\ref{fig:neutrino_flux} illustrates the predicted flux for each neutrino flavor and its parent hadrons from several generators at the FPF. 
For electron neutrinos, semileptonic kaon decays dominate up to approximately 1 TeV, above which charmed hadrons ($D$, $D_s$, $\Lambda_c$) become the primary source. 
Hyperons also make a smaller but notable contribution to the $\nu_e$ flux. 
On the other hand, muon neutrinos receive an additional significant component from pion decay, which dominates up to several hundred GeV. 
The tau neutrino flux originates almost entirely from charmed hadron decays.
Note that {\tt DPMJET} (dashed lines in the figure) has recently been known to be unreliable for such predictions, and is now also ruled out by data (see Fig.~9 of Ref.~\cite{FASER:2024ref}; see also the appendix of Ref.~\cite{FASER:2024ykc} for a detailed comparison and explanation).
See also Ref.~\cite{FASER:2024ykc} for refined flux predictions and description of uncertainties.

Two features merit particular attention. 
First, the $\nu_e$:$\nu_\mu$:$\nu_\tau$ flux ratio is roughly 0.1:1:$10^{-3}$. 
In fact, the numbers of neutrino events expected at the FPF neutrino detectors are about $10^5$ $\nu_e$ events, $10^6$ $\nu_\mu$ events, and somewhere between $2-20\times 10^3$ $\nu_\tau$ interactions, depending on the generator prediction for the total HL-LHC luminosity of 3~ab$^{-1}$.
Second, while predictions from different generators vary by at most a factor of two for neutrinos from pions, kaons, and hyperons, the charmed hadron production shows discrepancies spanning approximately one order of magnitude. 
Although this variation appears concerning, only one of the generators shown here, {\tt SIBYLL 2.3d}, has been calibrated with forward charm production data.
This highlights the need for dedicated studies of charm production in other generators, and for the most recent efforts, see, e.g., Ref.~\cite{Buonocore:2023kna} (charm production prediction from {\tt POWHEG} + {\tt PYTHIA}) and Refs.~\cite{Bai:2020ukz, Maciula:2022lzk, Bhattacharya:2023zei}.

Understanding the sources of the large discrepancies in charm production is crucial.
The uncertainties in branching ratios of $D$ and $D_s$ mesons typically remain below 10\%, thus contributing minimally to the variation among generators. Furthermore, since the branching ratio $D \to \tau\nu_\tau$ ($\sim$0.1\%) is approximately 50 times smaller than $D_s \to \tau\nu_\tau$ ($\sim$5\%), $D$ meson production has a negligible impact on the tau neutrino flux. Consequently, we can focus our discussion on uncertainties stemming from $D_s$ production. The predominant source of uncertainty in charm production stems from our limited knowledge of PDFs and hadronization processes.

For concreteness, we focus on the FPF, for which PDFs are dominated by the high-energy, forward-direction regime where even low momentum fraction $x$ can yield significant production cross sections for heavy mesons. This phenomenon is readily understood by examining the partonic center-of-mass energy in a proton-proton collision: $\hat{s} = x_1 x_2 s$, where $x_1$ and $x_2$ represent the momentum fractions carried by the incoming partons in each proton. 
PDFs at small $x$ values present measurement challenges, and the scarcity of experimental data in this regime leads to substantial uncertainties. In the very forward region, variations in renormalization and factorization scales can result in nearly an order-of-magnitude spread in predicted $D_s$ production at the FPF. This situation underscores the need for more precise theoretical estimates of $D$-meson production in the forward region, incorporating higher-order corrections, as well as forthcoming experimental data from LHCb on $D$-meson production.

\section{Neutrino Interactions at Colliders}

\begin{figure}[h]
	\centering
    \includegraphics[width=\columnwidth]{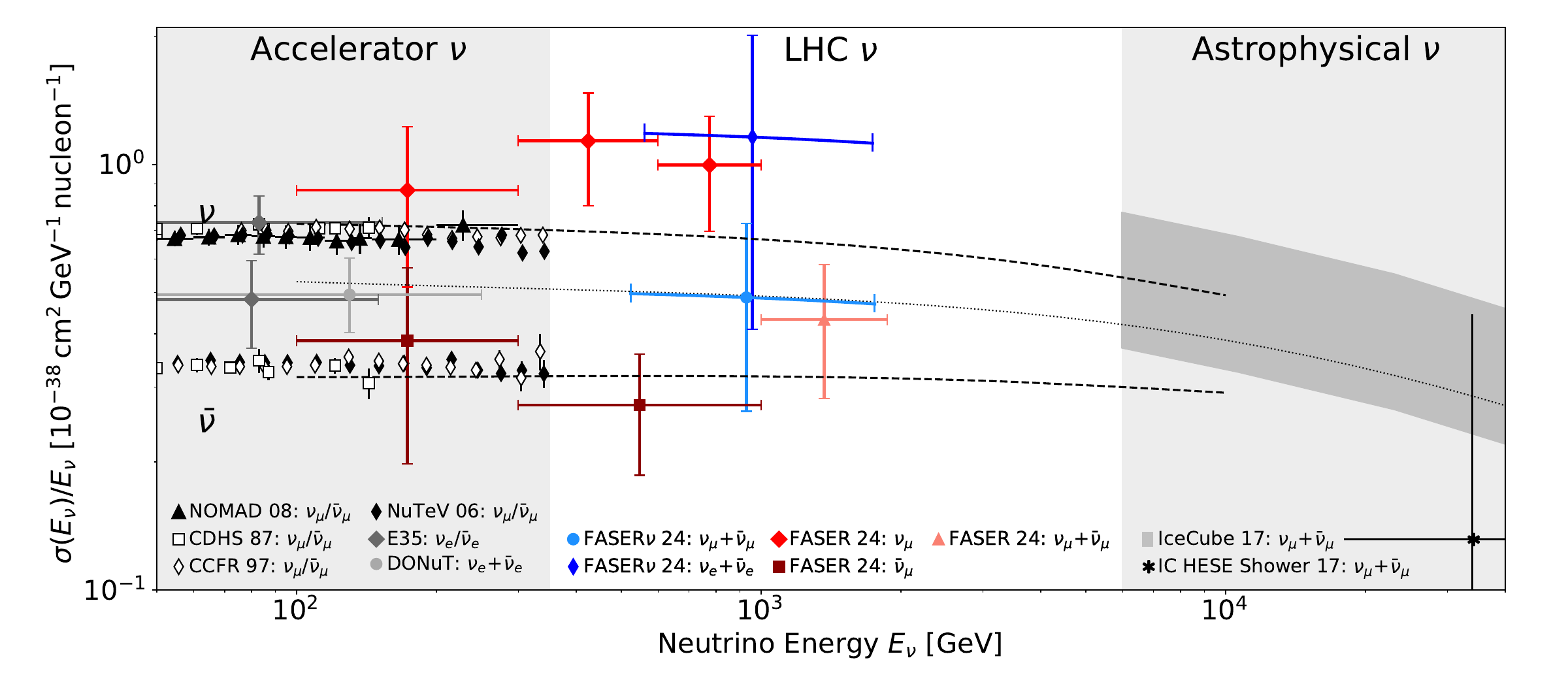}
	\caption{Neutrino cross section measurements above tens GeV, shown as points with error, with the corresponding experiments in the legend. The dashed contours correspond to the cross sections predicted by the Bodek-Yang model, as implemented in {\tt GENIE}. Note that the displayed experiments do not all use the same targets. Figure directly from Ref.~\cite{Ariga:2025qup}, which is a combination of Fig.~4 in Ref.~\cite{FASER:2024hoe} and Fig.~3 in Ref.~\cite{FASER:2024ref}.}
	\label{Fig_nu_xsec_measured}
\end{figure}

Fig.~\ref{Fig_nu_xsec_measured} shows the neutrino cross section measurements from 50~GeV to 40~TeV~\cite{Ariga:2025qup}.
Before FASER$\nu$, there were no measurements of the neutrino cross sections in the energy range from about 400 GeV to 6 TeV.
Above about 6~TeV, the cross sections have been measured by IceCube using Earth absorption effects~\cite{IceCube:2017roe, Bustamante:2017xuy, IceCube:2020rnc}.
Below about 400~GeV, the cross sections have been measured using accelerator neutrinos from, e.g., NuTeV~\cite{NuTeV:2005wsg}, CCFR~\cite{Seligman:1997fe}, and NOMAD~\cite{NOMAD:2007krq}.
Collider neutrinos will be crucial to fill the gap, with potential discoveries like neutrino-nucleus trident production~\cite{Kozhushner1961, Shabalin1963, Czyz:1964zz, Lovseth:1971vv, Fujikawa:1971nx, Koike:1971tu, Koike:1971vg, Brown:1972vne, Belusevic:1987cw, Altmannshofer:2014pba, Magill:2016hgc, Ge:2017poy, Ballett:2018uuc, Altmannshofer:2019zhy, Gauld:2019pgt, Zhou:2019vxt, Zhou:2019frk, Altmannshofer:2024hqd, Bigaran:2024zxk, Francener:2024wul} and W-boson production~\cite{Seckel:1997kk, Alikhanov:2015kla, Zhou:2019vxt, Zhou:2019frk, Xie:2023qbn}. 

In this section, we first review the neutrino interaction processes relevant to collider neutrinos (Sec.~\ref{sec_nuInt_processes}). Then, we discuss the existing measurements from FASER$\nu$ and SND@LHC and their prospects for the remaining LHC Run 3 (Sec.~\ref{sec_nuInt_existing}). Finally, we discuss the prospects for FPF (Sec.~\ref{sec_nuInt_FPF}).

\subsection{Interactions of Collider Neutrinos}
\label{sec_nuInt_processes}

Most of the collider neutrino events are from DIS, which dominates neutrino cross sections for $E_\nu$ above tens of GeV. 
The DIS can be either CC, i.e., 
$ \nu_{\ell} (\bar{\nu}_{\ell}) +A \rightarrow \ell^{-} (\ell^{+}) + X $, which is mediated by a W boson, or NC, i.e., $\stackrel{(-)}{\nu_{\ell}}+A \rightarrow \stackrel{(-)}{\nu_{\ell}}+X$, which is mediated by a Z boson. Here and below, $A$ and $A'$ denote initial and final-state nuclei, and $X$ denotes all the final-state hadrons.
The weak bosons couple to a quark in a nucleon in a nucleus ($A$).
The cross sections per nucleon at 100~GeV are $\sim 5 \times 10^{-37}$~cm$^2$ for CCDIS and $\sim 2 \times 10^{-37}$~cm$^2$ for NCDIS. Above 100~GeV, they increase linearly on $E_\nu$ and then the increase slows down and finally saturates at $E_\nu^{0.3}$ above $\sim 10^6$~GeV, due to the masses of the weak bosons ($\simeq 80$~GeV). See, e.g., Refs.~\cite{Ansari:2021cao, Candido:2023utz, Xie:2023suk, Jeong:2023hwe, Weigel:2024gzh, FerrarioRavasio:2024kem, vanBeekveld:2024ziz} for the most recent calculations of the neutrino DIS.

Despite the collider neutrinos from the LHC typically having energies above 100~GeV, the energies transferred to the nuclei in the detectors can occasionally be much smaller, which leads to QES and RES, also mediated by weak bosons. 
The QES can be described by $ \nu_{\ell} (\bar{\nu}_{\ell}) +A \rightarrow \ell^{-} (\ell^{+}) + A' $ (CC) and $\stackrel{(-)}{\nu_{\ell}}+A \rightarrow \stackrel{(-)}{\nu_{\ell}}+A$ (NC). The RES is similar, except that a nucleon in the nucleus is excited to a higher-energy state, e.g., $\Delta(1232)$ baryon, which usually decays to a proton (or neutron) and a pion.

There are other interesting neutrino interactions that are mediated by photon coupling to the nucleus, including neutrino-nucleus trident production~\cite{Kozhushner1961, Shabalin1963, Czyz:1964zz, Lovseth:1971vv, Fujikawa:1971nx, Koike:1971tu, Koike:1971vg, Brown:1972vne, Belusevic:1987cw, Altmannshofer:2014pba, Magill:2016hgc, Ge:2017poy, Ballett:2018uuc, Altmannshofer:2019zhy, Gauld:2019pgt, Zhou:2019vxt, Zhou:2019frk, Altmannshofer:2024hqd, Bigaran:2024zxk, Francener:2024wul} and W-boson production (WBP)~\cite{Seckel:1997kk, Alikhanov:2015kla, Zhou:2019vxt, Zhou:2019frk, Xie:2023qbn},
which have never been discovered at the 5$\sigma$ level and could be discovered with collider neutrinos~\cite{Altmannshofer:2024hqd}. 
The WBP is $\nu_\ell (\bar{\nu}_\ell) + A \rightarrow \ell^- (\ell^+) + W^+(W^-) + A'$. Due to the heavy $W$ mass, this process has a threshold of $E_\nu \sim$ TeV. 
The trident production is a three-lepton production process, i.e., $\stackrel{(-)}{\nu} + A \rightarrow \ell^- + \stackrel{(-)}{\nu}  + \ell^+ + A'$ (the diagrams of tridents in the Four-Fermi theory look like tridents; see, e.g., Fig.~2 of Ref.~\cite{Zhou:2019vxt}). 
WBP and tridents can happen in three different scattering regimes~\cite{Zhou:2019vxt}: 
1) coherent regime, in which the photon is so soft that it couples to the Coulomb field of the whole nucleus, which is analogous to the coherent neutrino-nucleus scattering~\cite{COHERENT:2017ipa}, 
2) diffractive regime, in which the photon couples to a nucleon, which is analogous to the QES,
and 3) inelastic regime, in which the photon couples to a quark, which is analogous to DIS. The sum of the cross sections in the three regimes gives the total cross sections of WBP and tridents.
At GeV energies, neutrino trident cross sections are only $\sim 10^{-4}$ of the CC cross section, making them extremely hard to detect. Above $\sim 10$~TeV energies, tridents are a part of WBP~\cite{Zhou:2019vxt}. Detecting these processes is crucial for testing BSM physics~\cite{Altmannshofer:2014pba, Magill:2016hgc, Ge:2017poy, Altmannshofer:2019zhy, Altmannshofer:2024hqd}.
In addition, resonance production of light mesons from $\bar{\nu}_e+e^-$ scattering is also relevant for collider neutrinos~\cite{Brdar:2021hpy}.
In particular, Ref.~\cite{Brdar:2021hpy} predicts that FASER$\nu$2 can collect about 30 events from $\bar{\nu}_e + e^- \rightarrow \rho^-$, which can potentially be identified from $\rho^- \rightarrow \pi^- \pi^0$.

Last but not least, the corrections from final-state radiation (FSR) to the CCDIS [i.e., $\nu_{\ell} (\bar{\nu}_{\ell}) +A \rightarrow \ell^{-} (\ell^{+}) + X + \gamma$] may be relevant for FASER$\nu$ and SND@LHC and will be relevant for the experiments at FPF~\cite{Plestid:2024bva}. FSR mainly impacts the differential cross sections, as the emitted photon takes away energy from the final-state charged lepton and is typically indistinguishable from the hadronic cascade, which increases the cascade energy. At 1~TeV, the reduction in the charged lepton energy is about 2\% and the increase in the cascade energy is about 4\%, which is important considering the large statistics of collider neutrino events. For PDF measurements, the distortion in the measured Bjorken-$x$ and factorization scale $Q$ due to FSR can be much larger.

\subsection{Existing Results at the LHC and Prospects for the Remaining Run 3}
\label{sec_nuInt_existing}

Fig.~\ref{Fig_nu_xsec_measured} shows the existing measurements of neutrino interaction cross sections using collider neutrinos~\cite{Ariga:2025qup}.
The two blue points are from the first measurement of high-energy $\nu_e$ and $\nu_\mu$ CC interactions in the FASER$\nu$ emulsion-tungsten detector of the FASER experiment at the LHC in March 2024~\cite{FASER:2024hoe}.
The measurement uses a 128.8 kg subset of the FASER$\nu$ volume and an exposure of 9.5 fb$^{-1}$of $\sqrt{s} = 13.6$~TeV $p$ $p$ data.
In total, 4 $\nu_e$ and 8 $\nu_\mu$ interaction candidate events are observed with a statistical significance of 5.2$\sigma$ and 5.7$\sigma$, respectively.
Based on the events observed, they made the first measurements of the neutrino cross section (per nucleon) in the energy range of $560-1740$~GeV for $\nu_e$ and $520-1760$~GeV for $\nu_\mu$, which gives $\sigma / E_\nu$ of $1.2_{-0.7}^{+0.8} \times 10^{-38} \mathrm{~cm}^2 \, \mathrm{GeV}^{-1}$ and $0.5 \pm 0.2 \times 10^{-38} \mathrm{~cm}^2 \, \mathrm{GeV}^{-1}$, respectively, consistent with the Standard Model predictions. 
The red points are from a more recent measurement using the electronic detector with 1.1~ton target mass in FASER, which managed to separate neutrinos and antineutrinos using muon appearance and muon-momentum measurements with the FASER spectrometer~\cite{FASER:2023zcr, FASER:2024ref}.

FASER$\nu$ and SND@LHC will continue taking data for the remaining period of LHC Run 3. Meanwhile, the understanding of the detector performance is improving. These will lead not only to more precise measurements of neutrino interactions but also to new detections, including NC interactions, $\nu_\tau$ processes, charm production, and new BSM searches. 
In addition, by matching the $\stackrel{(-)}{\nu}_\mu$ CC events in the emulsion detector to the electronic detector, the muons' momentum and charge can be measured by the spectrometer, which would allow separate measurements of neutrinos and antineutrinos.
Finally, a third veto station was recently installed at SND@LHC, and the entire veto system was improved. This will significantly increase the statistics of neutrino interactions in SND@LHC.

\subsection{Prospects for FPF}
\label{sec_nuInt_FPF}
Compared with FASER$\nu$ and SND running at the LHC, FPF will be running at the HL-LHC with much bigger detectors (see, e.g., table~1 of Ref.~\cite{Altmannshofer:2024hqd} about the detector comparisons). As a result, FPF will collect $\mathcal{O}(10^6)$ neutrino interaction events~\cite{Feng:2022inv}, much more than $\mathcal{O}(10^4)$ from FASER$\nu$ and SND. This corresponds to an order-of-magnitude improvement in the precision of the neutrino cross-section measurement. These not only apply to the dominant DIS interactions, but also QES, RES, tridents, and so on.

Measuring these interactions more precisely will also significantly enhance the precision of PDF measurements~\cite{Cruz-Martinez:2023sdv}, including the nuclear uncertainty, which will be beneficial for collider physics, high- and ultrahigh-energy neutrino astrophysics~\cite{Ackermann:2022rqc}, and other applications. 
In particular, the neutrino dimuon production from DIS has been an important process for measuring the strange-quark PDF (e.g., Refs.~\cite {DeLellis:2004ovi, Hou:2019efy, Faura:2020oom, Zhou:2021xuh}).
So far, only dimuon data from accelerator neutrino experiments are available for PDF fitting, and the ($x$, $Q$) coverage is only about (0.01--0.5, 1--10~GeV) (see, e.g., Fig.~9 of Ref.~\cite{Zhou:2021xuh}).
According to DIS kinematics, $Q^2 \simeq 4 E_\nu E_\mu \sin ^2\left(\theta_{\mu \mu} / 2\right)$ and $x \simeq$ $4 E_\nu E_\mu \sin ^2\left(\theta_{\mu \mu} / 2\right) /\left(2 m_N E_h\right)$, where $\theta_{\mu\mu}$ is the angular separation between the two outgoing muons, $m_N$ the nucleon mass, and $E_h$ the final-state hadronic energy. 
Therefore, $Q_{\max} \sim \operatorname{maximum} E_\nu$, $Q_{\min } \sim E_{\mathrm{th}} \theta_{\mu \mu}^{\min}$, and $x_{\min } \sim E_{\mathrm{th}}\left(\theta_{\mu \mu}^{\min }\right)^2$, where $E_{\mathrm{th}}$ and $\theta_{\mu \mu}^{\min}$ are the energy and angular thresholds of the detector. Thus, since FPF can measure neutrino interactions at energies one order of magnitude higher than accelerator neutrino experiments (Fig.~\ref{Fig_nu_xsec_measured}), the $Q_{\rm max}$ it can cover will be $\sim 3$ times higher. Moreover, with much lower energy and angular thresholds, FPF can also cover much smaller $x$ and $Q$ values. Importantly, lowering the angular threshold only moderately would significantly decrease $x_{\min }$, as $x_{\min } \sim \left(\theta_{\mu \mu}^{\min }\right)^2$.

FPF will also very likely make the first discovery of neutrino trident production~\cite{Altmannshofer:2024hqd}. The recent calculations from Refs.~\cite{Altmannshofer:2024hqd, Bigaran:2024zxk, Francener:2024wul} show that FASER$\nu2$ can collect about 40 $\mu^+\mu^-$, 44 $e^+ e^-$, 0.5 $\tau^+\tau^-$, 150 $e \mu$, 6 $e \tau$, and 10 $\mu \tau$ trident events (see, e.g, table I of Ref.~\cite{Altmannshofer:2024hqd} for more detectors and details). Importantly, Ref.~\cite{Altmannshofer:2024hqd} proposed a ``reverse tracking'' strategy and performed the first and detailed study of all possible backgrounds for trident detection and found that the $\mu^+\mu^-$ trident can be detected at FASER$\nu2$ at about $10\sigma$.

\section{Beyond the Standard Model Physics from Collider Neutrinos}

Collider neutrino experiments host a vibrant program to search for BSM physics~\cite{FASER:2018eoc, Feng:2022inv, Cheung:2023gwm}.
The unique experimental configuration---with detectors placed hundreds of meters downstream from the collision point---provides exceptional sensitivity to light, weakly-coupled particles that may be missed in central detectors.

The reason why such an experimental setup is special is a combination of high center-of-mass energy at the LHC and the location of the detector in the forward region. 
The forward region, or in collider terms, the large pseudo-rapidity, can probe collisions at very low momentum fraction (small-$x$).
This leads to substantially larger production rates of, for example, light and rare mesons, as the gluon PDF grows rapidly as $x$ decreases.
For reference, during the HL-LHC era, the FPF will see approximately $4\times10^{17}$ neutral pions, $6\times10^{16}$ eta mesons, $2\times10^{15}$ D mesons, and $1\times10^{13}$ B mesons produced in the far-forward region.

The sensitivity to small $x$ PDFs makes FPF particularly sensitive to light BSM particles produced either via meson decays, as well as via bremsstrahlung and Drell-Yan production mechanisms, which scale favorably with increasing center-of-mass energy.
We focus on theoretically well-motivated ``portal'' models: vector, fermion, and scalar portals.
Note that FPF can actually probe a much larger portfolio of BSM scenarios than what is presented here.
Nevertheless, these portal models represent minimal extensions to the Standard Model where new particles interact via mixing with known particles.
One more framework is worth mentioning due to its motivation and simplicity: axion-like particles (ALPs).
Many of these scenarios can also address outstanding questions of the standard model. 
Kinetic mixing scenarios are commonly found in dark matter models, the fermion portal is a natural consequence of neutrino mass models, axions can be related to the strong CP problem, and so on.
In what follows, we will describe the basic ideas behind each of these BSM scenarios and discuss how FPF can search for them.

In vector portal models, a new $U(1)$ gauge boson, typically referred to as a ``dark photon,'' kinetically mixes with the Standard Model photon through a dimensionless parameter $\varepsilon$.
The Lagrangian term for this mixing is
\begin{equation}
  \mathcal{L}\supset -\frac{\varepsilon}{2} F^{\mu\nu} F'_{\mu\nu},
\end{equation}
where $F$ and $F'$ are the field strength tensors of the photon $A$ and the dark photon $A'$, respectively.
By redefining  $A\to A + \varepsilon A'$, we can obtain properly diagonal and normalized kinetic terms for these fields to first order in $\varepsilon$.
This redefinition induces couplings between the dark photon and the electromagnetic current, namely $\varepsilon J_{\rm em}^\mu A'_\mu$: for every electromagnetic interaction in the standard model there is now a corresponding dark photon coupling.
This allows dark photons to be produced through processes similar to photon production, such as $\pi^0 \to \gamma A'$, $\eta \to \gamma A'$, and other meson decays; while the typical decays involve charged lepton pairs $\ell^+\ell^-$ or hadrons.
See, e.g., Ref.~\cite{FASER:2023tle} for a recent search for dark photons with FASER.

The fermion portal introduces new fermions that mix with Standard Model particles.
For light fermions, below the weak scale, the only experimentally viable option involves mixing with neutrinos, $\mathcal{L}\supset -y\overline L \tilde H N$, where $N$ are typically called heavy neutral leptons, or HNLs for short.
HNLs appear naturally in seesaw neutrino mass models, and may play a role in generating the matter-antimatter asymmetry of the universe via the leptogenesis mechanism.
These particles can be produced through charged meson decays due to their mixing with neutrinos, e.g., $\pi^+ \to \mu^+ N$ or $K^+ \to e^+ N$.
The same mixing drives their decay via weak interactions.

Scalar portals feature new light scalar particles, $\phi$, that mix with the Higgs boson through terms like $\mathcal{L}\supset\kappa|H|^2\phi$ in the scalar potential.
These scalars can be related to spontaneous symmetry breaking of new symmetries, such as a dark gauged $U(1)$, and may accompany, for example, models of dark photons and dark matter.
This mixing grants the new scalar Yukawa-like couplings to Standard Model fermions, with couplings proportional to the fermion masses.
Consequently, these scalars preferentially couple to heavy quarks when kinematically accessible, which will affect their production rates.
These scalars can be produced through meson decays like $K \to \pi\phi$ or $B \to K\phi$, or via gluon fusion like the Higgs; they typically decay to the heaviest available fermion pairs.

Last but not least, ALPs represent another compelling class of BSM candidates.
These pseudoscalar particles can couple to photons, fermions, or gluons through dimension-5 operators like
\begin{equation}
  \mathcal{L} \supset \frac{a}{f_a}F^{\mu\nu} \tilde{F}_{\mu\nu} + \frac{\partial_{\mu}a}{f_a} \bar{f}\gamma^{\mu}f + \frac{a}{f_a}G^{\mu\nu} \tilde{G}_{\mu\nu},
\end{equation}
where $a$ is the ALP, $f$ denotes a standard model fermion, and $G$ is the gluon field strength.
ALPs arise naturally in many beyond Standard Model theories, particularly as a solution to why $CP$ is apparently conserved in strong interactions.
They can be produced, again, through meson decays, as well as through their interactions with photons (e.g., Primakoff production). 
See, e.g., Ref.~\cite{FASER:2024bbl} for a recent search for ALPs with FASER.

For all these scenarios, when the BSM particle is light and weakly coupled, it can be long-lived.
They would be produced at or near the interaction point in the LHC central detectors, travel hundreds of meters, and decay within the volume of FPF detectors.
Therefore, experiments must distinguish potential BSM particle decay signals from the substantial background, which is mostly comprised of neutrino interactions.
Fortunately, the event topologies predicted in these BSM scenarios can be substantially different from neutrino events, for example, $e^+e^-$, $\mu^+\mu^-$, or specific hadronic final states; as opposed to usual DIS events.
The FPF detectors can separate the BSM signatures from backgrounds through precision tracking, calorimetry, and particle identification techniques, by leveraging distinct kinematic features---such as vertex displacement, invariant mass peaks, or characteristic angular distributions.

\begin{figure}[h]
\centering
\begin{minipage}[b]{0.47\columnwidth}
    \centering
    \includegraphics[width=\columnwidth]{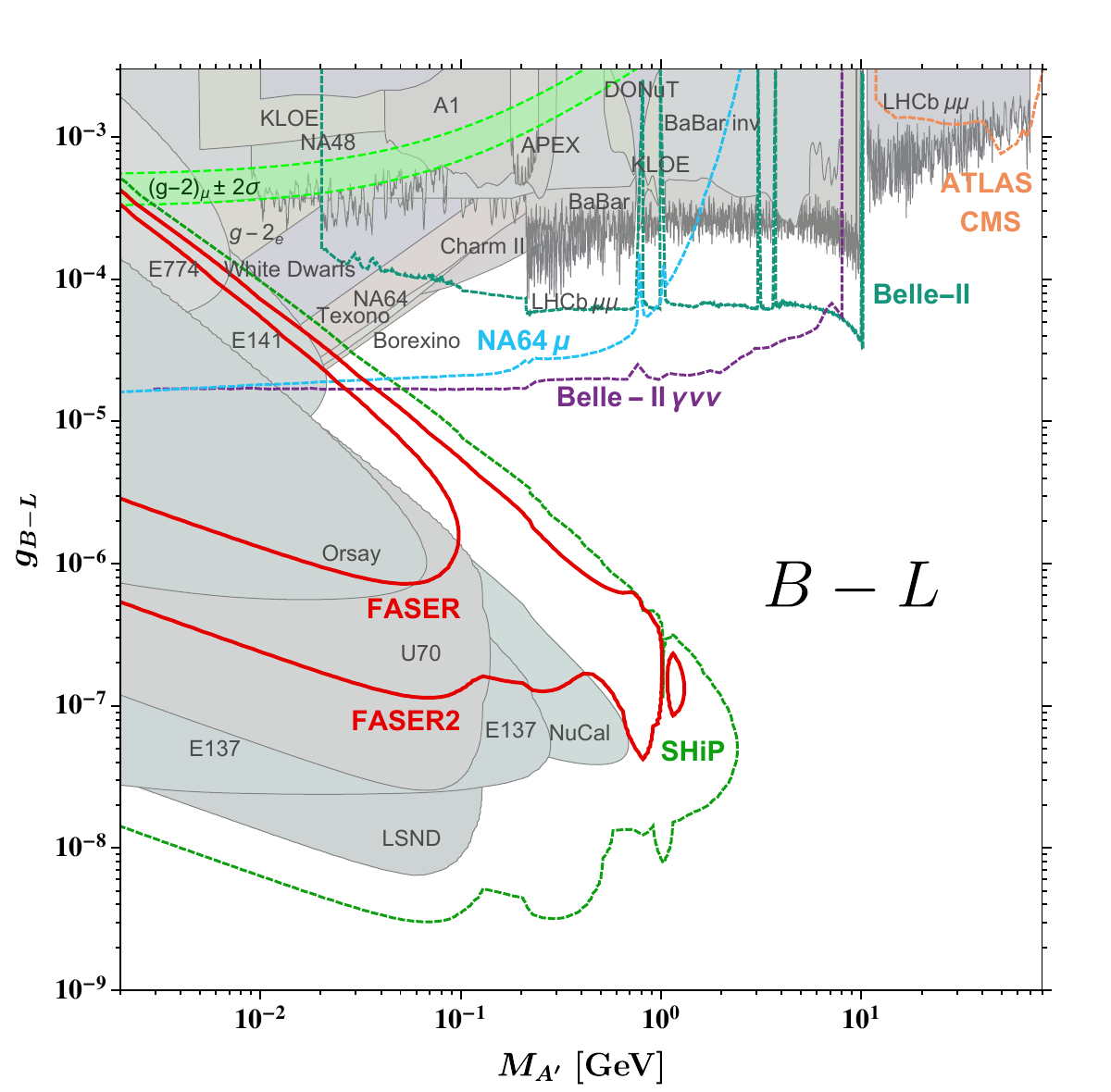}
\end{minipage}\hfill
\begin{minipage}[b]{0.49\columnwidth}
    \centering
    \includegraphics[width=0.935\columnwidth]{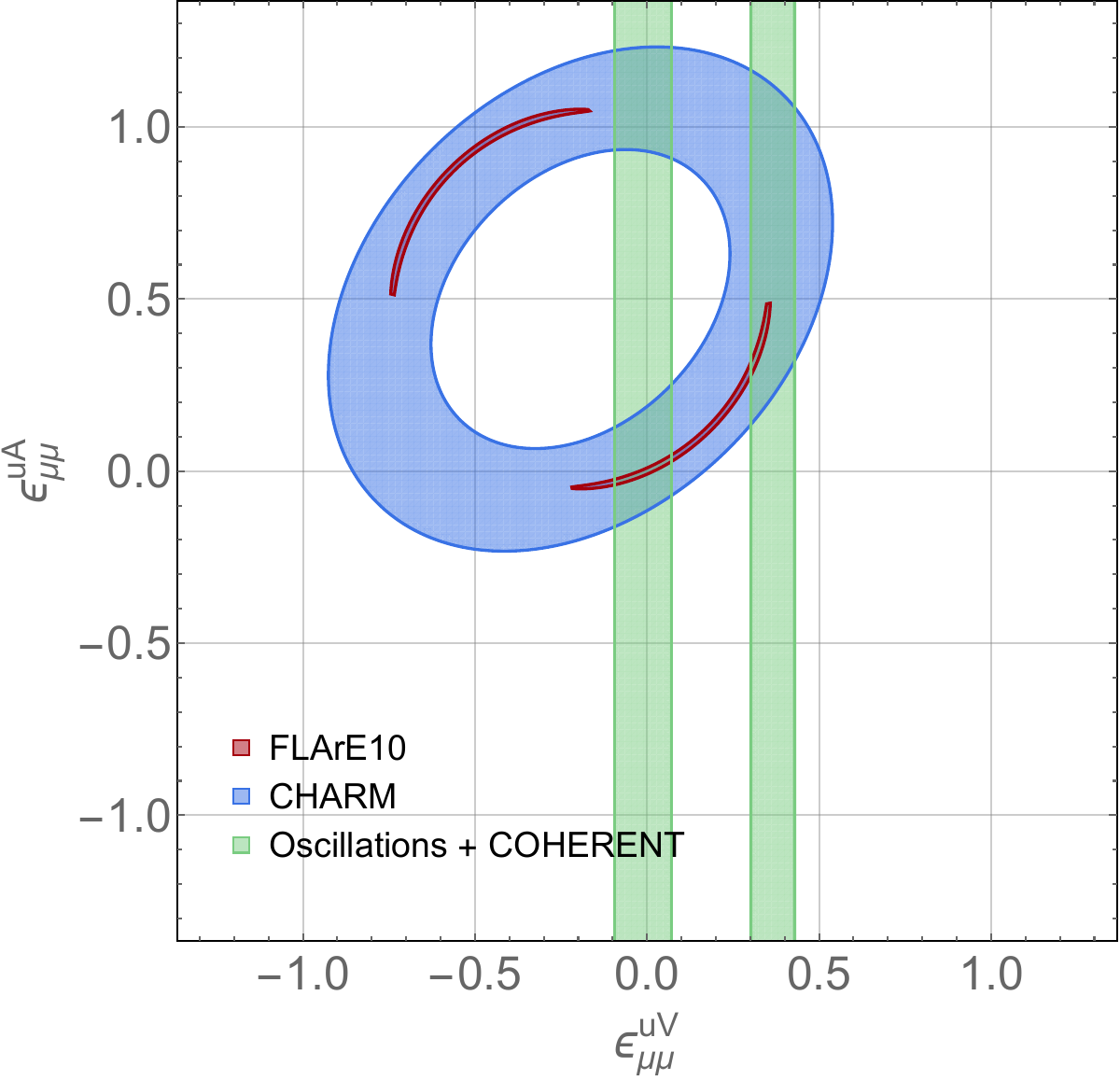}
    \vspace{-0.15cm}
\end{minipage}
\caption{Left: Projected sensitivity of FASER and FASER2 experiments to a light $U(1)_{B-L}$ gauge boson in the plane of coupling mass $M_{A'}$ versus gauge coupling $g_{B-L}$. Right: FLArE sensitivity to neutrino nonstandard interactions. Figures taken from Ref.~\cite{Feng:2022inv}.
}
\label{fig:bsm_sensitivity}
\end{figure}

Beyond these portal scenarios, the FPF also provides unique sensitivity to BSM physics within the neutrino sector.
Two simple possibilities are sterile neutrinos and non-standard interactions.
Sterile neutrinos are essentially HNLs, though typically are light enough that they may induce oscillation physics.
When a neutrino of a given flavor is produced, it may contain a small admixture of a fourth, mostly sterile state.
With a mass-squared difference between neutrino mass eigenstates $\Delta m_{41}^2\equiv m_4^2-m_1^2 \sim 10^3$ eV$^2$, and an energy $E \sim 1$ TeV, sterile neutrinos could induce oscillations at baselines of $L \sim 600$ m, leading to nontrivial flavor conversions at the FPF.
Given the substantial asymmetry between muon and tau neutrino fluxes in the forward region, FPF would be in an excellent position to probe $\nu_\mu \to \nu_\tau$ appearance due to sterile neutrinos by observing an excess of $\nu_\tau$ events over expectations. In addition, sterile neutrinos also distort the energy spectrum of the observed neutrinos at FPF~\cite{MammenAbraham:2024gun}.

Neutrino non-standard interactions (NSIs) represent another promising target.
These NSIs are effective four-fermion interactions between neutrinos and usual matter.
They can be described by effective field theory operators like
\begin{equation}
  \mathcal{L}\supset 2\sqrt{2}G_F\varepsilon^f_{\alpha\beta}(\bar{\nu}_\alpha \gamma^\mu \nu_\beta)(\bar{f} \gamma_\mu f)
                   + 2\sqrt{2}G_F\overline\varepsilon^f_{\alpha\beta}(\bar{\nu}_\alpha \gamma^\mu \ell_\beta)(\bar{f} \gamma_\mu f'),
\end{equation}
where $\alpha$ and $\beta$ are flavor indices and $f,f'$ represent a standard model fermions.
The first operator describes NC NSIs while the second is referred to as CC NSIs. 
Note that while these interactions are effective, written below electroweak symmetry breaking, they must come from a proper ultraviolet complete scenario involving new degrees of freedom either below or above the electroweak scale. 
Regardless, such interactions could, for example, induce processes like $\nu_\mu + N \to \nu_\tau + X$, which are not present in the standard model.
The FPF's high-statistics neutrino sample can be used to set strong constraints on many of these operators. For detailed studies, see, e.g., Ref.~\cite{Ismail:2020yqc} on neutral-current NSIs and Refs.~\cite{Falkowski:2021bkq, Kling:2023tgr} on charged-current NSIs.

This short overview captures only a small fraction of the BSM physics potential at the FPF.
To give a flavor of the sensitivities of collider neutrinos to new physics, we present representative cases in Fig.~\ref{fig:bsm_sensitivity}.  
Additional scenarios include inelastic dark matter models, strongly interacting massive particles, millicharged particles, and various non-minimal extensions of the dark sector with multiple new particles and interactions. See, e.g., Ref.~\cite{Ariga:2025qup} for a brief discussion of probing with muons collected by the same detectors.
The FPF's unique combination of high energy, intense neutrino flux, and long baseline provides complementary sensitivity to many BSM scenarios that would be challenging to probe with other experimental approaches.

\section{Astrophysics from Collider Neutrinos}
\label{sec_astroph}

Neutrinos from colliders are also crucial for astrophysical neutrino studies, as collider neutrino data will improve our understanding of the production and interaction of neutrinos, which are essential for astrophysical observations. Collider neutrinos will be another bridge between particle physics and astrophysics.

\subsection{Understanding Atmospheric Neutrino Background for High-Energy Astrophysical Neutrino Observations}

The observation of high-energy astrophysical neutrinos by IceCube in 2013~\cite{IceCube:2013low} has opened a new window into astrophysics and particle physics~\cite{Ackermann:2022rqc}.
These neutrinos are essentially the only tool for studying the origin of high-energy cosmic rays --- a century-long puzzle. Moreover, they are also the primary tool for studying dense and other extreme astrophysical environments, e.g., choked-jet supernovae~\cite{Senno:2015tsn, Senno:2017vtd, Esmaili:2018wnv, Chang:2022hqj, IceCube:2023esf}. Last but not least, these neutrinos have the unique power to test a lot of BSM physics~\cite{Ackermann:2022rqc}.

However, atmospheric neutrinos~\cite{Gaisser:2002jj} are an irreducible background for observing astrophysical high-energy neutrinos and searching for and studying their sources (see, e.g., Fig.~2 of Ref.~\cite{Chang:2022hqj}).
These neutrinos originate from the semileptonic decays of hadrons produced by cosmic rays bombarding the Earth's atmosphere nuclei and can be categorized into conventional neutrinos and prompt neutrinos. The former comes from the decays of light hadrons, including $\pi^{\pm}$ and $K^{\pm}$. The latter comes from the decays of heavier hadrons, including $D$ mesons, $B$ mesons, and $A_c$ baryons.

The conventional atmospheric neutrino flux falls steeply with increasing energy~\cite{Gaisser:2002jj}. Its spectral shape reflects that of the incoming cosmic ray flux ($\sim E^{-2.7}$, where $E$ is the energy) plus an additional suppression ($\sim E^{-1}$) due to the energy losses experienced by the mesons ($\pi^{\pm}$, $K^{\pm}$) before decay.

The prompt neutrino flux is increasingly important at higher energies. Unlike $\pi^{\pm}$ and $K^{\pm}$, which are responsible for the conventional flux, the heavy hadrons responsible for the prompt flux decay immediately to neutrinos after they are produced because of their much shorter decay length. 
Therefore, there is no additional suppression of $\sim E^{-1}$ in the prompt flux, making them more important at higher energies. It is predicted that the prompt neutrinos start to dominate over the conventional neutrinos around $E_\nu = 10^5$--$10^6$~GeV. 
Although prompt atmospheric neutrinos have not yet been observed, their flux uncertainties are now one of the most important systematic uncertainty sources for measurements of high-energy astrophysical neutrinos at neutrino observatories such as IceCube~\cite{IceCube_web}, KM3NeT~\cite{KM3Net:2016zxf}, and Baikal-GVD~\cite{Allakhverdyan:2021vkk}. 
Prompt atmospheric neutrinos are even more important for the future IceCube-Gen2~\cite{IceCube-Gen2:2020qha, Gen2_TDR_web}, P-ONE~\cite{P-ONE:2020ljt}, TRIDENT~\cite{Ye:2023dch}, HUNT~\cite{Huang:2023mzt}, and NEON~\cite{Zhang:2024slv}, which will observe neutrinos at energies up to $\sim 10$ times higher.

Therefore, neutrino astronomy necessitates an in-depth understanding of these atmospheric neutrinos. Collider neutrinos, especially FPF, will provide key information on this. 
For the conventional neutrino flux, the FPF will provide data to test and tune the Monte Carlo codes in the forward region in various ways~\cite{Fieg:2023kld}. These measurements can check individual perturbative Quantum Chromodynamics calculations and help to reduce their associated uncertainties. Moreover, these data can serve as input for theoretical calculations, thus improving existing models. 

For prompt neutrino production during cosmic ray interactions in the atmosphere, heavy flavors are predominantly produced in the forward direction, as the partons involved, primarily gluons, have very asymmetric longitudinal momentum fractions.
For example, for cosmic-ray proton energy of $E_p = 100$~TeV, the momentum fractions of the two incoming gluons are $x_1\simeq1$ and $x_2\simeq3\times10^{-5}$. For $E_p = 1$~PeV, $x_1\simeq1$ and $x_2\simeq3\times10^{-6}$.
Therefore, these processes are characterized by very small Bjorken-$x$ values.

The FPF is instrumental in enhancing our understanding and measurement of prompt neutrino flux~\cite{Feng:2022inv}. These facilities can investigate the small-$x$ region discussed above and also cover a wide range of rapidity, which is crucial for accurately modeling heavy-flavor production and the resultant neutrino flux. For instance, charm mesons (the dominant hadrons that produce prompt neutrinos)
produced in the forward region of cosmic ray interactions have been measured in the rapidity range of $2.0 < y < 4.5$ at the LHCb, while the forward experiments during Run 3 of the LHC will cover $y > 7.2$.
By extending the rapidity range to include more forward regions, FPF will provide critical data that can improve the precision in the predictions of prompt atmospheric neutrino fluxes, which will significantly help with the study of high-energy astrophysical neutrinos.

Last but not least, the improved modeling will also improve the understanding of small-scale features in the neutrino spectra. Possible examples include the accurate detection of the cosmic ray knee in the atmospheric neutrino spectrum, as well as a more precise description of the impact of seasonal modulations in atmospheric temperature on the spectra.

\subsection{Improving Understanding of Cosmic Ray Air Showers}

Cosmic rays comprise highly energetic particles arriving at Earth from astrophysical sources. 
Although observed up to energies of $\sim 10^{11}\,\mathrm{GeV}$, 
their acceleration mechanisms and nuclear composition remain only partially understood. 
Most of the information about cosmic rays above ${\sim}10^6\,\mathrm{GeV}$ 
comes from extensive air showers (EAS), 
where the primary cosmic ray interacts with atmospheric nuclei, 
producing a cascade of secondary particles that reach ground‐based detectors. 
The uncertainties in the EAS modeling dominate the uncertainties in the measurements of the UHE CR compositions, which limit our understanding of the acceleration mechanism.
One of the main challenges of the EAS modeling is the persistent 
\emph{muon puzzle}: current hadronic interaction models 
systematically underestimate the number of muons measured at the ground, and the deviation increases with energy. This discrepancy was first reported over two decades ago, and it persists across multiple experiments.

The so-called $z$-scale is defined to make muon measurements in different experiments comparable, i.e.,
\begin{equation}
z=\frac{\ln \left\langle N^{\rm obs}_\mu\right\rangle-\ln \left\langle N^{\rm pred}_{\mu, p}\right\rangle}{\ln \left\langle N^{\rm pred}_{\mu, \mathrm{Fe}}\right\rangle-\ln \left\langle N^{\rm pred}_{\mu, p}\right\rangle},
\end{equation}
where $\left\langle N^{\rm obs}_\mu\right\rangle$ is the observed average muon density, and $\left\langle N^{\rm pred}_{\mu, p}\right\rangle$ and $\left\langle N^{\rm pred}_{\mu, \mathrm{Fe}}\right\rangle$ are the predicted average muon densities for proton and iron showers, respectively, in a full detector simulation. 
Thus, \(z=0\) would imply perfect agreement for a proton‐initiated shower, 
whereas \(z=1\) would match an iron‐initiated shower. 
The $z$-scale, after an energy cross-calibration, is approximately independent of the experimental details while primarily depending on the hadronic interaction modeling.
The difference between experimental observation ($z_{\rm obs}$) and theoretical predictions ($z_{\rm pred}$; taking into account the mass composition dependence), $\Delta z (E) \equiv z_{\rm obs} - z_{\rm pred}$, was expected to be consistent with zero. However, it is observed to linearly increase with the primary energy $E$, and the slope of this increase deviates from zero with a significance of $>8\sigma$~\cite{Soldin:2021wyv}.

Several models are proposed to explain such deviations~\cite{Farrar:2013sfa, Anchordoqui:2016oxy, Baur:2019cpv, Anchordoqui:2019laz, Pierog:2020ghc}. 
A common feature of these models is that the neutral particle production is suppressed compared to the current models (e.g. 
{\tt EPOS-LHC}~\cite{Pierog:2013ria}, {\tt QGSJet-II.04}~\cite{Ostapchenko:2013pia}, {\tt Sibyll-2.3c/d}~\cite{Riehn:2019jet, Riehn:2017mfm}, and {\tt DPMJet-III.2017}~\cite{Roesler:2000he, Fedynitch:2115393}. 
The consistent increase of $\Delta z$ with energy favors a small suppression over a large energy range rather than a strong suppression occurring in the first few interactions of the EAS.

The FPF will play a crucial role in understanding the muon puzzle, in particular, in the amount of forward strangeness production as a consequence of the above proposed explanations~\cite{Anchordoqui:2022fpn, Feng:2022inv}.
The forward strangeness production is characterized by the ratio of charged kaons to pions, which can be determined through the ratio of electron and muon neutrino fluxes at the FPF. The electron fluxes are a proxy of kaon production, and both electron and muon neutrino fluxes are proxies of pion production. In addition, the kinematic distributions are different in the neutrinos from pion and kaon decays, and in electron and muon neutrinos from pion decays. Utilizing this information, FPF can measure the pion-to-kaon ratio. 
Another possible way for FPF to measure the pion-to-kaon ratio is to use the forward-going muons. During FASER run 3 (2022--24), about $10^9$ muons were detected, and FASER2 in the FPF is expected to detect $\sim 1000$ times more. These high-energy neutrinos and muons from colliders will complement those from IceCube and other telescopes, helping to reduce the systematic uncertainties in IceCube's measurements.

\section{Conclusions}
\label{sec:conclusions}

Collider neutrino experiments are opening a vibrant new window into neutrino physics and astrophysics, bridging collider-based particle physics, astrophysics, and BSM physics. The pioneering efforts of experiments like FASER$\nu$ and SND@LHC have already provided groundbreaking measurements of neutrino cross sections at previously inaccessible energies. The upcoming FPF at the HL-LHC promises significant advancements, from precision cross-section measurements and improved proton structure insights to potential discoveries of new physics including dark matter candidates, sterile neutrinos, and non-standard neutrino interactions.
The impacts of these collider neutrino experiments extend beyond particle physics into astrophysics. 
They will significantly enhance our understanding of conventional and prompt atmospheric neutrino fluxes, one of the largest sources of systematic uncertainties at neutrino observatories like IceCube and KM3NeT. In addition, they will help solve the long-standing “muon puzzle” problem in cosmic-ray measurements.
Looking forward, further experimental and theoretical efforts at collider facilities, particularly at the FPF, will continue to sharpen our understanding of neutrino properties, cosmic ray interactions, and the fundamental forces shaping our universe.

\begin{ack}[Acknowledgments]
We thank Jamie Boyd, Jonathan Feng, Max Fieg, and Felix Kling for helpful comments on the draft manuscript.
This work is supported by Fermi Forward Discovery Group, LLC under Contract No. 89243024CSC000002 with the U.S. Department of Energy, Office of Science, Office of High Energy Physics.

\end{ack}


\bibliographystyle{Numbered-Style}
\bibliography{reference}

\begin{thebibliography*}{100}
\providecommand{\bibtype}[1]{}
\providecommand{\url}[1]{{\tt #1}}
\providecommand{\urlprefix}{URL }
\expandafter\ifx\csname urlstyle\endcsname\relax
  \providecommand{\doi}[1]{doi:\discretionary{}{}{}#1}\else
  \providecommand{\doi}{doi:\discretionary{}{}{}\begingroup
  \urlstyle{rm}\Url}\fi
\providecommand{\bibinfo}[2]{#2}
\providecommand{\eprint}[2][]{\url{#2}}
\makeatletter\def\@biblabel#1{\bibinfo{label}{[#1]}}\makeatother

\bibtype{Article}%
\bibitem{FASER:2019dxq}
\bibinfo{author}{Henso Abreu}, et al. (\bibinfo{collaboration}{FASER}),
  \bibinfo{title}{{Detecting and Studying High-Energy Collider Neutrinos with
  FASER at the LHC}}, \bibinfo{journal}{Eur. Phys. J. C} \bibinfo{volume}{80}
  (\bibinfo{number}{1}) (\bibinfo{year}{2020}) \bibinfo{pages}{61},
  \bibinfo{doi}{\doi{10.1140/epjc/s10052-020-7631-5}}, \eprint{1908.02310}.

\bibtype{Article}%
\bibitem{FASER:2020gpr}
\bibinfo{author}{Henso Abreu}, et al. (\bibinfo{collaboration}{FASER}),
  \bibinfo{title}{{Technical Proposal: FASERnu}}  (\bibinfo{year}{2020}),
  \eprint{2001.03073}.

\bibtype{Article}%
\bibitem{FASER:2023zcr}
\bibinfo{author}{Henso Abreu}, et al. (\bibinfo{collaboration}{FASER}),
  \bibinfo{title}{{First Direct Observation of Collider Neutrinos with FASER at
  the LHC}}  (\bibinfo{year}{2023}), \eprint{2303.14185}.

\bibtype{Article}%
\bibitem{Feng:2022inv}
\bibinfo{author}{Jonathan~L. Feng}, et al., \bibinfo{title}{{The Forward
  Physics Facility at the High-Luminosity LHC}}, \bibinfo{journal}{J. Phys. G}
  \bibinfo{volume}{50} (\bibinfo{number}{3}) (\bibinfo{year}{2023})
  \bibinfo{pages}{030501}, \bibinfo{doi}{\doi{10.1088/1361-6471/ac865e}},
  \eprint{2203.05090}.

\bibtype{Misc}%
\bibitem{FPF_web}
\bibinfo{howpublished}{\url{https://fpf.web.cern.ch/}}.

\bibtype{Article}%
\bibitem{CCFR:1997tam}
\bibinfo{author}{W.~G. Seligman}, et al. (\bibinfo{collaboration}{CCFR}),
  \bibinfo{title}{{Improved determination of $\alpha_s$ from neutrino nucleon
  scattering}}, \bibinfo{journal}{Phys. Rev. Lett.} \bibinfo{volume}{79}
  (\bibinfo{year}{1997}) \bibinfo{pages}{1213--1216},
  \bibinfo{doi}{\doi{10.1103/PhysRevLett.79.1213}}, \eprint{hep-ex/9701017}.

\bibtype{Article}%
\bibitem{NuTeV:2003kth}
\bibinfo{author}{D. Naples}, et al. (\bibinfo{collaboration}{NuTeV}),
  \bibinfo{title}{{High energy neutrino scattering results from NuTeV}},
  \bibinfo{journal}{Nucl. Phys. B Proc. Suppl.} \bibinfo{volume}{118}
  (\bibinfo{year}{2003}) \bibinfo{pages}{164--173},
  \bibinfo{doi}{\doi{10.1016/S0920-5632(03)01314-8}}.

\bibtype{Article}%
\bibitem{NOMAD:2009qmu}
\bibinfo{author}{V Lyubushkin}, et al. (\bibinfo{collaboration}{NOMAD}),
  \bibinfo{title}{{A Study of quasi-elastic muon neutrino and antineutrino
  scattering in the NOMAD experiment}}, \bibinfo{journal}{Eur. Phys. J. C}
  \bibinfo{volume}{63} (\bibinfo{year}{2009}) \bibinfo{pages}{355--381},
  \bibinfo{doi}{\doi{10.1140/epjc/s10052-009-1113-0}}, \eprint{0812.4543}.

\bibtype{Article}%
\bibitem{Jain:2000pu}
\bibinfo{author}{P. Jain}, \bibinfo{author}{Douglas~W. McKay},
  \bibinfo{author}{S. Panda}, \bibinfo{author}{John~P. Ralston},
  \bibinfo{title}{{Extra dimensions and strong neutrino nucleon interactions
  above 10**19-eV: Breaking the GZK barrier}}, \bibinfo{journal}{Phys. Lett. B}
  \bibinfo{volume}{484} (\bibinfo{year}{2000}) \bibinfo{pages}{267--274},
  \bibinfo{doi}{\doi{10.1016/S0370-2693(00)00647-X}}, \eprint{hep-ph/0001031}.

\bibtype{Article}%
\bibitem{Arguelles:2015wba}
\bibinfo{author}{Carlos~A. Arg\"uelles}, \bibinfo{author}{Francis Halzen},
  \bibinfo{author}{Logan Wille}, \bibinfo{author}{Mike Kroll},
  \bibinfo{author}{Mary~Hall Reno}, \bibinfo{title}{{High-energy behavior of
  photon, neutrino, and proton cross sections}}, \bibinfo{journal}{Phys. Rev.
  D} \bibinfo{volume}{92} (\bibinfo{number}{7}) (\bibinfo{year}{2015})
  \bibinfo{pages}{074040}, \bibinfo{doi}{\doi{10.1103/PhysRevD.92.074040}},
  \eprint{1504.06639}.

\bibtype{Article}%
\bibitem{Becirevic:2018uab}
\bibinfo{author}{Damir Be\v{c}irevi\'c}, \bibinfo{author}{Boris Panes},
  \bibinfo{author}{Olcyr Sumensari}, \bibinfo{author}{Renata
  Zukanovich~Funchal}, \bibinfo{title}{{Seeking leptoquarks in IceCube}},
  \bibinfo{journal}{JHEP} \bibinfo{volume}{06} (\bibinfo{year}{2018})
  \bibinfo{pages}{032}, \bibinfo{doi}{\doi{10.1007/JHEP06(2018)032}},
  \eprint{1803.10112}.

\bibtype{Article}%
\bibitem{Bai:2025pef}
\bibinfo{author}{Yang Bai}, \bibinfo{author}{Keping Xie}, \bibinfo{author}{Bei
  Zhou}, \bibinfo{title}{{Large Neutrino ''Collider''}}
  (\bibinfo{year}{2025}), \eprint{2510.13948}.

\bibtype{Article}%
\bibitem{Ariga:2025qup}
\bibinfo{author}{Akitaka Ariga}, \bibinfo{author}{Jamie Boyd},
  \bibinfo{author}{Felix Kling}, \bibinfo{author}{Albert De~Roeck},
  \bibinfo{title}{{Neutrino Experiments at the Large Hadron Collider}}
  (\bibinfo{year}{2025}),
  \bibinfo{doi}{\doi{10.1146/annurev-nucl-121423-101000}}, \eprint{2501.10078}.

\bibtype{Article}%
\bibitem{Amoroso:2022eow}
\bibinfo{author}{S. Amoroso}, et al., \bibinfo{title}{{Snowmass 2021
  Whitepaper: Proton Structure at the Precision Frontier}},
  \bibinfo{journal}{Acta Phys. Polon. B} \bibinfo{volume}{53}
  (\bibinfo{number}{12}) (\bibinfo{year}{2022}) \bibinfo{pages}{12--A1},
  \bibinfo{doi}{\doi{10.5506/APhysPolB.53.12-A1}}, \eprint{2203.13923}.

\bibtype{Article}%
\bibitem{Cruz-Martinez:2023sdv}
\bibinfo{author}{Juan~M. Cruz-Martinez}, \bibinfo{author}{Max Fieg},
  \bibinfo{author}{Tommaso Giani}, \bibinfo{author}{Peter Krack},
  \bibinfo{author}{Toni M\"akel\"a}, \bibinfo{author}{Tanjona~R.
  Rabemananjara}, \bibinfo{author}{Juan Rojo}, \bibinfo{title}{{The LHC as a
  Neutrino-Ion Collider}}, \bibinfo{journal}{Eur. Phys. J. C}
  \bibinfo{volume}{84} (\bibinfo{number}{4}) (\bibinfo{year}{2024})
  \bibinfo{pages}{369}, \bibinfo{doi}{\doi{10.1140/epjc/s10052-024-12665-1}},
  \eprint{2309.09581}.

\bibtype{Article}%
\bibitem{Albrecht:2021cxw}
\bibinfo{author}{Johannes Albrecht}, et al., \bibinfo{title}{{The Muon Puzzle
  in cosmic-ray induced air showers and its connection to the Large Hadron
  Collider}}, \bibinfo{journal}{Astrophys. Space Sci.} \bibinfo{volume}{367}
  (\bibinfo{number}{3}) (\bibinfo{year}{2022}) \bibinfo{pages}{27},
  \bibinfo{doi}{\doi{10.1007/s10509-022-04054-5}}, \eprint{2105.06148}.

\bibtype{Article}%
\bibitem{FASER:2025qaf}
\bibinfo{author}{Roshan Mammen~Abraham}, et al.
  (\bibinfo{collaboration}{FASER}), \bibinfo{title}{{Reconstruction and
  Performance Evaluation of FASER's Emulsion Detector at the LHC}}
  (\bibinfo{year}{2025}), \eprint{2504.13008}.

\bibtype{Article}%
\bibitem{FASER:2022hcn}
\bibinfo{author}{Henso Abreu}, et al. (\bibinfo{collaboration}{FASER}),
  \bibinfo{title}{{The FASER detector}}, \bibinfo{journal}{JINST}
  \bibinfo{volume}{19} (\bibinfo{number}{05}) (\bibinfo{year}{2024})
  \bibinfo{pages}{P05066}, \bibinfo{doi}{\doi{10.1088/1748-0221/19/05/P05066}},
  \eprint{2207.11427}.

\bibtype{Article}%
\bibitem{Arakawa:2022rmp}
\bibinfo{author}{Jason Arakawa}, \bibinfo{author}{Jonathan~L. Feng},
  \bibinfo{author}{Ahmed Ismail}, \bibinfo{author}{Felix Kling},
  \bibinfo{author}{Michael Waterbury}, \bibinfo{title}{{Neutrino detection
  without neutrino detectors: Discovering collider neutrinos at FASER with
  electronic signals only}}, \bibinfo{journal}{Phys. Rev. D}
  \bibinfo{volume}{106} (\bibinfo{number}{5}) (\bibinfo{year}{2022})
  \bibinfo{pages}{052011}, \bibinfo{doi}{\doi{10.1103/PhysRevD.106.052011}},
  \eprint{2206.09932}.

\bibtype{Article}%
\bibitem{Feng:2017uoz}
\bibinfo{author}{Jonathan~L. Feng}, \bibinfo{author}{Iftah Galon},
  \bibinfo{author}{Felix Kling}, \bibinfo{author}{Sebastian Trojanowski},
  \bibinfo{title}{{ForwArd Search ExpeRiment at the LHC}},
  \bibinfo{journal}{Phys. Rev. D} \bibinfo{volume}{97} (\bibinfo{number}{3})
  (\bibinfo{year}{2018}) \bibinfo{pages}{035001},
  \bibinfo{doi}{\doi{10.1103/PhysRevD.97.035001}}, \eprint{1708.09389}.

\bibtype{Article}%
\bibitem{FASER:2018bac}
\bibinfo{author}{Akitaka Ariga}, et al. (\bibinfo{collaboration}{FASER}),
  \bibinfo{title}{{Technical Proposal for FASER: ForwArd Search ExpeRiment at
  the LHC}}  (\bibinfo{year}{2018}), \eprint{1812.09139}.

\bibtype{Article}%
\bibitem{FASER:2018ceo}
\bibinfo{author}{Akitaka Ariga}, et al. (\bibinfo{collaboration}{FASER}),
  \bibinfo{title}{{Letter of Intent for FASER: ForwArd Search ExpeRiment at the
  LHC}}  (\bibinfo{year}{2018}), \eprint{1811.10243}.

\bibtype{Article}%
\bibitem{SNDLHC:2022ihg}
\bibinfo{author}{G. Acampora}, et al. (\bibinfo{collaboration}{SND@LHC}),
  \bibinfo{title}{{SND@LHC: the scattering and neutrino detector at the LHC}},
  \bibinfo{journal}{JINST} \bibinfo{volume}{19} (\bibinfo{number}{05})
  (\bibinfo{year}{2024}) \bibinfo{pages}{P05067},
  \bibinfo{doi}{\doi{10.1088/1748-0221/19/05/P05067}}, \eprint{2210.02784}.

\bibtype{Misc}%
\bibitem{SND_web}
\bibinfo{howpublished}{\url{https://snd-lhc.web.cern.ch/}}.

\bibtype{Techreport}%
\bibitem{Abbaneo:2926288}
\bibinfo{author}{D Abbaneo}, \bibinfo{author}{C Ahdida}, \bibinfo{author}{S
  Ahmad}, \bibinfo{author}{R Albanese}, \bibinfo{author}{A Alexandrov},
  \bibinfo{author}{F Alicante}, \bibinfo{author}{F Aloschi}, \bibinfo{author}{N
  Amapane}, \bibinfo{author}{M Andreini}, \bibinfo{author}{K Androsov},
  \bibinfo{author}{A Anokhina}, \bibinfo{author}{T Asada}, \bibinfo{author}{C
  Asawatangtrakuldee}, \bibinfo{author}{M~A Ayala~Torres}, \bibinfo{author}{N
  Bangaru}, \bibinfo{author}{C Battilana}, \bibinfo{author}{A Bay},
  \bibinfo{author}{A Bertocco}, \bibinfo{author}{C Bertone}, \bibinfo{author}{C
  Betancourt}, \bibinfo{author}{D Bick}, \bibinfo{author}{R Biswas},
  \bibinfo{author}{A Blanco~Castro}, \bibinfo{author}{V Boccia},
  \bibinfo{author}{O Boettcher}, \bibinfo{author}{M Bogomilov},
  \bibinfo{author}{D Bonacorsi}, \bibinfo{author}{W~M Bonivento},
  \bibinfo{author}{P Bordalo}, \bibinfo{author}{A Boyarsky},
  \bibinfo{author}{T~A Bud}, \bibinfo{author}{L Buonocore}, \bibinfo{author}{S
  Buontempo}, \bibinfo{author}{V Cafaro}, \bibinfo{author}{T Camporesi},
  \bibinfo{author}{V Canale}, \bibinfo{author}{D Centanni}, \bibinfo{author}{F
  Cerutti}, \bibinfo{author}{A Cervelli}, \bibinfo{author}{V Chariton},
  \bibinfo{author}{N Charitonidis}, \bibinfo{author}{M Chernyavskiy},
  \bibinfo{author}{A Chiuchiolo}, \bibinfo{author}{K~Y Choi},
  \bibinfo{author}{S Cholak}, \bibinfo{author}{F Cindolo}, \bibinfo{author}{M
  Climescu}, \bibinfo{author}{A~P Conaboy}, \bibinfo{author}{O Crespo~Lopez},
  \bibinfo{author}{A Crupano}, \bibinfo{author}{D D'Agostino},
  \bibinfo{author}{G~M Dallavalle}, \bibinfo{author}{N D'Ambrosio},
  \bibinfo{author}{D Davino}, \bibinfo{author}{R De~Asmundis},
  \bibinfo{author}{P~T De~Bryas}, \bibinfo{author}{G De~Lellis},
  \bibinfo{author}{M De~Magistris}, \bibinfo{author}{G De~Marzi},
  \bibinfo{author}{S De~Pasquale}, \bibinfo{author}{A De~Roeck},
  \bibinfo{author}{A De~Rujula}, \bibinfo{author}{D De~Simone},
  \bibinfo{author}{A Di~Crescenzo}, \bibinfo{author}{D Di~Ferdinando},
  \bibinfo{author}{L Di~Giulio}, \bibinfo{author}{S Di~Luca},
  \bibinfo{author}{C Dinc}, \bibinfo{author}{R Don\`a}, \bibinfo{author}{O
  Durhan}, \bibinfo{author}{D Fasanella}, \bibinfo{author}{M Ferrillo},
  \bibinfo{author}{R~A Fini}, \bibinfo{author}{A Fiorillo}, \bibinfo{author}{R
  Fresa}, \bibinfo{author}{W Funk}, \bibinfo{author}{N Funicello},
  \bibinfo{author}{C Gaignant}, \bibinfo{author}{V Giordano},
  \bibinfo{author}{A Golutvin}, \bibinfo{author}{E Graverini},
  \bibinfo{author}{L Guiducci}, \bibinfo{author}{A~M Guler}, \bibinfo{author}{V
  Guliaeva}, \bibinfo{author}{G~J Haefeli}, \bibinfo{author}{C Hagner},
  \bibinfo{author}{J~C Helo~Herrera}, \bibinfo{author}{A Herty},
  \bibinfo{author}{E Van~Herwijnen}, \bibinfo{author}{A Iaiunese},
  \bibinfo{author}{S Ilieva}, \bibinfo{author}{A Infantino}, \bibinfo{author}{A
  Iuliano}, \bibinfo{author}{H Jeangros}, \bibinfo{author}{C Kamiscioglu},
  \bibinfo{author}{A~M Kauniskangas}, \bibinfo{author}{S~H Kim},
  \bibinfo{author}{Y~G Kim}, \bibinfo{author}{G Klioutchnikov},
  \bibinfo{author}{M Komatsu}, \bibinfo{author}{S Kuleshov}, \bibinfo{author}{L
  Krzempek}, \bibinfo{author}{H~M Lacker}, \bibinfo{author}{O Lantwin},
  \bibinfo{author}{F Lasagni~Manghi}, \bibinfo{author}{A Lauria},
  \bibinfo{author}{K~Y Lee}, \bibinfo{author}{K~S Lee}, \bibinfo{author}{P
  Lelong}, \bibinfo{author}{E Leo}, \bibinfo{author}{G Lerner},
  \bibinfo{author}{V~P Loschiavo}, \bibinfo{author}{G Magazzu},
  \bibinfo{author}{M Majstorovic}, \bibinfo{author}{S Marcellini},
  \bibinfo{author}{A Margiotta}, \bibinfo{author}{A~P Marion},
  \bibinfo{author}{A Mascellani}, \bibinfo{author}{F Mei}, \bibinfo{author}{A
  Miano}, \bibinfo{author}{A Mikulenko}, \bibinfo{author}{M~C Montesi},
  \bibinfo{author}{F~L Navarria}, \bibinfo{author}{E Nowak}, \bibinfo{author}{W
  Nuntiyakul}, \bibinfo{author}{S Ogawa}, \bibinfo{author}{J Osborne},
  \bibinfo{author}{M Ovchynnikov}, \bibinfo{author}{G Paggi},
  \bibinfo{author}{K Pal}, \bibinfo{author}{J Panigoni}, \bibinfo{author}{B~D
  Park}, \bibinfo{author}{S Pelletier}, \bibinfo{author}{M Perez~Ornedo},
  \bibinfo{author}{A Perrotta}, \bibinfo{author}{N Polukhina},
  \bibinfo{author}{F Primavera}, \bibinfo{author}{A Prota}, \bibinfo{author}{O
  Prouteau}, \bibinfo{author}{A Quercia}, \bibinfo{author}{S Ramos},
  \bibinfo{author}{A Reghunath}, \bibinfo{author}{F Ronchetti},
  \bibinfo{author}{T Rovelli}, \bibinfo{author}{O Ruchayskiy},
  \bibinfo{author}{M Sabate~Gilarte}, \bibinfo{author}{Z Sadykov},
  \bibinfo{author}{F Sanchez~Galan}, \bibinfo{author}{M Sarno},
  \bibinfo{author}{V Scalera}, \bibinfo{author}{W Schmidt-Parzefall},
  \bibinfo{author}{O Schneider}, \bibinfo{author}{G Sekhniaidze},
  \bibinfo{author}{N Serra}, \bibinfo{author}{M Shaposhnikov},
  \bibinfo{author}{T Shchedrina}, \bibinfo{author}{L Shchutska},
  \bibinfo{author}{H Shibuya}, \bibinfo{author}{A Sidoti}, \bibinfo{author}{G~P
  Siroli}, \bibinfo{author}{G Sirri}, \bibinfo{author}{G Soares},
  \bibinfo{author}{J~Y Sohn}, \bibinfo{author}{O~J Soto~Sandoval},
  \bibinfo{author}{J~L Soto~Pezoa}, \bibinfo{author}{M Spurio},
  \bibinfo{author}{J Steggemann}, \bibinfo{author}{M Szewczyk},
  \bibinfo{author}{I Timiryasov}, \bibinfo{author}{V Tioukov},
  \bibinfo{author}{M Tobar}, \bibinfo{author}{F Tramontano}, \bibinfo{author}{C
  Trippl}, \bibinfo{author}{A Uluwita}, \bibinfo{author}{E Ursov},
  \bibinfo{author}{G Vankova-Kirilova}, \bibinfo{author}{G Vasquez},
  \bibinfo{author}{V Verguilov}, \bibinfo{author}{N Viegas Guerreiro~Leonardo},
  \bibinfo{author}{C Vilela}, \bibinfo{author}{A Vieille}, \bibinfo{author}{C
  Visone}, \bibinfo{author}{R Wanke}, \bibinfo{author}{E Yaman},
  \bibinfo{author}{Z Yang}, \bibinfo{author}{E Yaman}, \bibinfo{author}{C
  Yazici}, \bibinfo{author}{C~S Yoon}, \bibinfo{author}{E Zaffaroni},
  \bibinfo{author}{J Zamora~Saa}, \bibinfo{author}{M Zanetti},
  \bibinfo{title}{{SND@HL-LHC, Scattering and Neutrino Detector in Run 4 of the
  LHC}}, \bibinfo{type}{\bibinfo{comment}{tech. rep.}},
  \bibinfo{institution}{CERN}, \bibinfo{address}{Geneva} \bibinfo{year}{2025},
  \bibinfo{url}{\urlprefix\url{https://cds.cern.ch/record/2926288}}.

\bibtype{Article}%
\bibitem{Anchordoqui:2021ghd}
\bibinfo{author}{Luis~A. Anchordoqui}, et al., \bibinfo{title}{{The Forward
  Physics Facility: Sites, experiments, and physics potential}},
  \bibinfo{journal}{Phys. Rept.} \bibinfo{volume}{968} (\bibinfo{year}{2022})
  \bibinfo{pages}{1--50}, \bibinfo{doi}{\doi{10.1016/j.physrep.2022.04.004}},
  \eprint{2109.10905}.

\bibtype{Article}%
\bibitem{Adhikary:2024nlv}
\bibinfo{author}{Jyotismita Adhikary}, et al., \bibinfo{title}{{Scientific
  program for the Forward Physics Facility}}, \bibinfo{journal}{Eur. Phys. J.
  C} \bibinfo{volume}{85} (\bibinfo{number}{4}) (\bibinfo{year}{2025})
  \bibinfo{pages}{430}, \bibinfo{doi}{\doi{10.1140/epjc/s10052-025-14048-6}},
  \eprint{2411.04175}.

\bibtype{Inproceedings}%
\bibitem{FPFWorkingGroups:2025rsc}
\bibinfo{author}{Luis~A. Anchordoqui}, et al. (\bibinfo{collaboration}{FPF
  Working Groups}), \bibinfo{title}{{The Forward Physics Facility at the Large
  Hadron Collider}} \bibinfo{year}{2025}, \eprint{2503.19010}.

\bibtype{Article}%
\bibitem{Kamp:2025phs}
\bibinfo{author}{Nicholas~W. Kamp}, \bibinfo{author}{Carlos~A. Arg{\"u}elles},
  \bibinfo{author}{Albrecht Karle}, \bibinfo{author}{Jennifer Thomas},
  \bibinfo{author}{Tianlu Yuan}, \bibinfo{title}{{Lake- and Surface-Based
  Detectors for Forward Neutrino Physics}}  (\bibinfo{year}{2025}),
  \eprint{2501.08278}.

\bibtype{Inproceedings}%
\bibitem{Roesler:2000he}
\bibinfo{author}{Stefan Roesler}, \bibinfo{author}{Ralph Engel},
  \bibinfo{author}{Johannes Ranft}, \bibinfo{title}{{The Monte Carlo event
  generator DPMJET-III}}, in: \bibinfo{booktitle}{{International Conference on
  Advanced Monte Carlo for Radiation Physics, Particle Transport Simulation and
  Applications (MC 2000)}} \bibinfo{year}{2000}, pp.
  \bibinfo{pages}{1033--1038},
  \bibinfo{doi}{\doi{10.1007/978-3-642-18211-2_166}}, \eprint{hep-ph/0012252}.

\bibtype{Article}%
\bibitem{Ahn:2009wx}
\bibinfo{author}{Eun-Joo Ahn}, \bibinfo{author}{Ralph Engel},
  \bibinfo{author}{Thomas~K. Gaisser}, \bibinfo{author}{Paolo Lipari},
  \bibinfo{author}{Todor Stanev}, \bibinfo{title}{{Cosmic ray interaction event
  generator SIBYLL 2.1}}, \bibinfo{journal}{Phys. Rev. D} \bibinfo{volume}{80}
  (\bibinfo{year}{2009}) \bibinfo{pages}{094003},
  \bibinfo{doi}{\doi{10.1103/PhysRevD.80.094003}}, \eprint{0906.4113}.

\bibtype{Article}%
\bibitem{Ostapchenko:2010vb}
\bibinfo{author}{Sergey Ostapchenko}, \bibinfo{title}{{Monte Carlo treatment of
  hadronic interactions in enhanced Pomeron scheme: I. QGSJET-II model}},
  \bibinfo{journal}{Phys. Rev. D} \bibinfo{volume}{83} (\bibinfo{year}{2011})
  \bibinfo{pages}{014018}, \bibinfo{doi}{\doi{10.1103/PhysRevD.83.014018}},
  \eprint{1010.1869}.

\bibtype{Article}%
\bibitem{Pierog:2013ria}
\bibinfo{author}{T. Pierog}, \bibinfo{author}{Iu. Karpenko},
  \bibinfo{author}{J.~M. Katzy}, \bibinfo{author}{E. Yatsenko},
  \bibinfo{author}{K. Werner}, \bibinfo{title}{{EPOS LHC: Test of collective
  hadronization with data measured at the CERN Large Hadron Collider}},
  \bibinfo{journal}{Phys. Rev. C} \bibinfo{volume}{92} (\bibinfo{number}{3})
  (\bibinfo{year}{2015}) \bibinfo{pages}{034906},
  \bibinfo{doi}{\doi{10.1103/PhysRevC.92.034906}}, \eprint{1306.0121}.

\bibtype{Article}%
\bibitem{Skands:2014pea}
\bibinfo{author}{Peter Skands}, \bibinfo{author}{Stefano Carrazza},
  \bibinfo{author}{Juan Rojo}, \bibinfo{title}{{Tuning PYTHIA 8.1: the Monash
  2013 Tune}}, \bibinfo{journal}{Eur. Phys. J. C} \bibinfo{volume}{74}
  (\bibinfo{number}{8}) (\bibinfo{year}{2014}) \bibinfo{pages}{3024},
  \bibinfo{doi}{\doi{10.1140/epjc/s10052-014-3024-y}}, \eprint{1404.5630}.

\bibtype{Article}%
\bibitem{Sjostrand:2014zea}
\bibinfo{author}{Torbj\"orn Sj\"ostrand}, \bibinfo{author}{Stefan Ask},
  \bibinfo{author}{Jesper~R. Christiansen}, \bibinfo{author}{Richard Corke},
  \bibinfo{author}{Nishita Desai}, \bibinfo{author}{Philip Ilten},
  \bibinfo{author}{Stephen Mrenna}, \bibinfo{author}{Stefan Prestel},
  \bibinfo{author}{Christine~O. Rasmussen}, \bibinfo{author}{Peter~Z. Skands},
  \bibinfo{title}{{An introduction to PYTHIA 8.2}}, \bibinfo{journal}{Comput.
  Phys. Commun.} \bibinfo{volume}{191} (\bibinfo{year}{2015})
  \bibinfo{pages}{159--177}, \bibinfo{doi}{\doi{10.1016/j.cpc.2015.01.024}},
  \eprint{1410.3012}.

\bibtype{Phdthesis}%
\bibitem{Fedynitch:2015kcn}
\bibinfo{author}{Anatoli Fedynitch}, \bibinfo{title}{{Cascade equations and
  hadronic interactions at very high energies}}, \bibinfo{comment}{Ph.D.
  thesis}, \bibinfo{school}{KIT, Karlsruhe, Dept. Phys.} \bibinfo{year}{2015},
  \bibinfo{doi}{\doi{10.5445/IR/1000055433}}.

\bibtype{Article}%
\bibitem{Riehn:2015oba}
\bibinfo{author}{Felix Riehn}, \bibinfo{author}{Ralph Engel},
  \bibinfo{author}{Anatoli Fedynitch}, \bibinfo{author}{Thomas~K. Gaisser},
  \bibinfo{author}{Todor Stanev}, \bibinfo{title}{{A new version of the event
  generator Sibyll}}, \bibinfo{journal}{PoS} \bibinfo{volume}{ICRC2015}
  (\bibinfo{year}{2016}) \bibinfo{pages}{558},
  \bibinfo{doi}{\doi{10.22323/1.236.0558}}, \eprint{1510.00568}.

\bibtype{Article}%
\bibitem{Riehn:2017mfm}
\bibinfo{author}{Felix Riehn}, \bibinfo{author}{Hans~P. Dembinski},
  \bibinfo{author}{Ralph Engel}, \bibinfo{author}{Anatoli Fedynitch},
  \bibinfo{author}{Thomas~K. Gaisser}, \bibinfo{author}{Todor Stanev},
  \bibinfo{title}{{The hadronic interaction model SIBYLL 2.3c and Feynman
  scaling}}, \bibinfo{journal}{PoS} \bibinfo{volume}{ICRC2017}
  (\bibinfo{year}{2018}) \bibinfo{pages}{301},
  \bibinfo{doi}{\doi{10.22323/1.301.0301}}, \eprint{1709.07227}.

\bibtype{Article}%
\bibitem{Fedynitch:2018cbl}
\bibinfo{author}{Anatoli Fedynitch}, \bibinfo{author}{Felix Riehn},
  \bibinfo{author}{Ralph Engel}, \bibinfo{author}{Thomas~K. Gaisser},
  \bibinfo{author}{Todor Stanev}, \bibinfo{title}{{Hadronic interaction model
  sibyll 2.3c and inclusive lepton fluxes}}, \bibinfo{journal}{Phys. Rev. D}
  \bibinfo{volume}{100} (\bibinfo{number}{10}) (\bibinfo{year}{2019})
  \bibinfo{pages}{103018}, \bibinfo{doi}{\doi{10.1103/PhysRevD.100.103018}},
  \eprint{1806.04140}.

\bibtype{Article}%
\bibitem{Battistoni:2015epi}
\bibinfo{author}{Giuseppe Battistoni}, et al., \bibinfo{title}{{Overview of the
  FLUKA code}}, \bibinfo{journal}{Annals Nucl. Energy} \bibinfo{volume}{82}
  (\bibinfo{year}{2015}) \bibinfo{pages}{10--18},
  \bibinfo{doi}{\doi{10.1016/j.anucene.2014.11.007}}.

\bibtype{Article}%
\bibitem{Nevay:2018zhp}
\bibinfo{author}{Laurence~J. Nevay}, et al., \bibinfo{title}{{BDSIM: An
  accelerator tracking code with particle\textendash{}matter interactions}},
  \bibinfo{journal}{Comput. Phys. Commun.} \bibinfo{volume}{252}
  (\bibinfo{year}{2020}) \bibinfo{pages}{107200},
  \bibinfo{doi}{\doi{10.1016/j.cpc.2020.107200}}, \eprint{1808.10745}.

\bibtype{Article}%
\bibitem{Kling:2021gos}
\bibinfo{author}{Felix Kling}, \bibinfo{author}{Laurence~J. Nevay},
  \bibinfo{title}{{Forward neutrino fluxes at the LHC}},
  \bibinfo{journal}{Phys. Rev. D} \bibinfo{volume}{104} (\bibinfo{number}{11})
  (\bibinfo{year}{2021}) \bibinfo{pages}{113008},
  \bibinfo{doi}{\doi{10.1103/PhysRevD.104.113008}}, \eprint{2105.08270}.

\bibtype{Article}%
\bibitem{FASER:2024ref}
\bibinfo{author}{Roshan Mammen~Abraham}, et al.
  (\bibinfo{collaboration}{FASER}), \bibinfo{title}{{First Measurement of the
  Muon Neutrino Interaction Cross Section and Flux as a Function of Energy at
  the LHC with FASER}}, \bibinfo{journal}{Phys. Rev. Lett.}
  \bibinfo{volume}{134} (\bibinfo{number}{21}) (\bibinfo{year}{2025})
  \bibinfo{pages}{211801}, \bibinfo{doi}{\doi{10.1103/PhysRevLett.134.211801}},
  \eprint{2412.03186}.

\bibtype{Article}%
\bibitem{FASER:2024ykc}
\bibinfo{author}{Roshan Mammen~Abraham}, et al.
  (\bibinfo{collaboration}{FASER}), \bibinfo{title}{{Neutrino Rate Predictions
  for FASER}}  (\bibinfo{year}{2024}), \eprint{2402.13318}.

\bibtype{Article}%
\bibitem{Buonocore:2023kna}
\bibinfo{author}{Luca Buonocore}, \bibinfo{author}{Felix Kling},
  \bibinfo{author}{Luca Rottoli}, \bibinfo{author}{Jonas Sominka},
  \bibinfo{title}{{Predictions for Neutrinos and New Physics from Forward Heavy
  Hadron Production at the LHC}}  (\bibinfo{year}{2023}), \eprint{2309.12793}.

\bibtype{Article}%
\bibitem{Bai:2020ukz}
\bibinfo{author}{Weidong Bai}, \bibinfo{author}{Milind Diwan},
  \bibinfo{author}{Maria~Vittoria Garzelli}, \bibinfo{author}{Yu~Seon Jeong},
  \bibinfo{author}{Mary~Hall Reno}, \bibinfo{title}{{Far-forward neutrinos at
  the Large Hadron Collider}}, \bibinfo{journal}{JHEP} \bibinfo{volume}{06}
  (\bibinfo{year}{2020}) \bibinfo{pages}{032},
  \bibinfo{doi}{\doi{10.1007/JHEP06(2020)032}}, \eprint{2002.03012}.

\bibtype{Article}%
\bibitem{Maciula:2022lzk}
\bibinfo{author}{Rafal Maciula}, \bibinfo{author}{Antoni Szczurek},
  \bibinfo{title}{{Far-forward production of charm mesons and neutrinos at
  forward physics facilities at the LHC and the intrinsic charm in the
  proton}}, \bibinfo{journal}{Phys. Rev. D} \bibinfo{volume}{107}
  (\bibinfo{number}{3}) (\bibinfo{year}{2023}) \bibinfo{pages}{034002},
  \bibinfo{doi}{\doi{10.1103/PhysRevD.107.034002}}, \eprint{2210.08890}.

\bibtype{Article}%
\bibitem{Bhattacharya:2023zei}
\bibinfo{author}{Atri Bhattacharya}, \bibinfo{author}{Felix Kling},
  \bibinfo{author}{Ina Sarcevic}, \bibinfo{author}{Anna~M. Stasto},
  \bibinfo{title}{{Forward Neutrinos from Charm at Large Hadron Collider}}
  (\bibinfo{year}{2023}), \eprint{2306.01578}.

\bibtype{Article}%
\bibitem{FASER:2024hoe}
\bibinfo{author}{Roshan Mammen~Abraham}, et al.
  (\bibinfo{collaboration}{FASER}), \bibinfo{title}{{First Measurement of
  \ensuremath{\nu}e and \ensuremath{\nu}\ensuremath{\mu} Interaction Cross
  Sections at the LHC with FASER\textquoteright{}s Emulsion Detector}},
  \bibinfo{journal}{Phys. Rev. Lett.} \bibinfo{volume}{133}
  (\bibinfo{number}{2}) (\bibinfo{year}{2024}) \bibinfo{pages}{021802},
  \bibinfo{doi}{\doi{10.1103/PhysRevLett.133.021802}}, \eprint{2403.12520}.

\bibtype{Article}%
\bibitem{IceCube:2017roe}
\bibinfo{author}{M.~G. Aartsen}, et al. (\bibinfo{collaboration}{IceCube}),
  \bibinfo{title}{{Measurement of the multi-TeV neutrino cross section with
  IceCube using Earth absorption}}, \bibinfo{journal}{Nature}
  \bibinfo{volume}{551} (\bibinfo{year}{2017}) \bibinfo{pages}{596--600},
  \bibinfo{doi}{\doi{10.1038/nature24459}}, \eprint{1711.08119}.

\bibtype{Article}%
\bibitem{Bustamante:2017xuy}
\bibinfo{author}{Mauricio Bustamante}, \bibinfo{author}{Amy Connolly},
  \bibinfo{title}{{Extracting the Energy-Dependent Neutrino-Nucleon Cross
  Section above 10 TeV Using IceCube Showers}}, \bibinfo{journal}{Phys. Rev.
  Lett.} \bibinfo{volume}{122} (\bibinfo{number}{4}) (\bibinfo{year}{2019})
  \bibinfo{pages}{041101}, \bibinfo{doi}{\doi{10.1103/PhysRevLett.122.041101}},
  \eprint{1711.11043}.

\bibtype{Article}%
\bibitem{IceCube:2020rnc}
\bibinfo{author}{R. Abbasi}, et al. (\bibinfo{collaboration}{IceCube}),
  \bibinfo{title}{{Measurement of the high-energy all-flavor neutrino-nucleon
  cross section with IceCube}}  (\bibinfo{year}{2020}),
  \bibinfo{doi}{\doi{10.1103/PhysRevD.104.022001}}, \eprint{2011.03560}.

\bibtype{Article}%
\bibitem{NuTeV:2005wsg}
\bibinfo{author}{M. Tzanov}, et al. (\bibinfo{collaboration}{NuTeV}),
  \bibinfo{title}{{Precise measurement of neutrino and anti-neutrino
  differential cross sections}}, \bibinfo{journal}{Phys. Rev. D}
  \bibinfo{volume}{74} (\bibinfo{year}{2006}) \bibinfo{pages}{012008},
  \bibinfo{doi}{\doi{10.1103/PhysRevD.74.012008}}, \eprint{hep-ex/0509010}.

\bibtype{Phdthesis}%
\bibitem{Seligman:1997fe}
\bibinfo{author}{William~Glenn Seligman}, \bibinfo{title}{{A Next-to-Leading
  Order QCD Analysis of Neutrino - Iron Structure Functions at the Tevatron}},
  \bibinfo{comment}{Ph.D. thesis}, \bibinfo{school}{Nevis Labs, Columbia U.}
  \bibinfo{year}{1997}, \bibinfo{doi}{\doi{10.2172/1421736}},
  \bibinfo{url}{\urlprefix\url{http://lss.fnal.gov/cgi-bin/find_paper.pl?thesis-1997-21}}.

\bibtype{Article}%
\bibitem{NOMAD:2007krq}
\bibinfo{author}{Q. Wu}, et al. (\bibinfo{collaboration}{NOMAD}),
  \bibinfo{title}{{A Precise measurement of the muon neutrino-nucleon inclusive
  charged current cross-section off an isoscalar target in the energy range 2.5
  \ensuremath{<} E(nu) \ensuremath{<} 40-GeV by NOMAD}},
  \bibinfo{journal}{Phys. Lett. B} \bibinfo{volume}{660} (\bibinfo{year}{2008})
  \bibinfo{pages}{19--25}, \bibinfo{doi}{\doi{10.1016/j.physletb.2007.12.027}},
  \eprint{0711.1183}.

\bibtype{Article}%
\bibitem{Kozhushner1961}
\bibinfo{author}{M.A. Kozhushner}, \bibinfo{author}{E.P. Shabalin},
  \bibinfo{title}{PRODUCTION OF LEPTON PARTICLE PAIRS ON A COULOMB CENTER},
  \bibinfo{journal}{Soviet Journal of Experimental and Theoretical Physics}
  \bibinfo{volume}{41} (\bibinfo{year}{1961}) \bibinfo{pages}{949}.

\bibtype{Article}%
\bibitem{Shabalin1963}
\bibinfo{author}{E.~P. {Shabalin}}, \bibinfo{title}{{The
  {$\mu$}$^{+}${$\mu$}$^{-}$ and e$^{+}$e$^{-}$ Pair Production Cross Sections
  for Neutrinos Scattered by Nuclei}}, \bibinfo{journal}{Soviet Journal of
  Experimental and Theoretical Physics} \bibinfo{volume}{16}
  (\bibinfo{year}{1963}) \bibinfo{pages}{125}.

\bibtype{Article}%
\bibitem{Czyz:1964zz}
\bibinfo{author}{W. Czyz}, \bibinfo{author}{G.~C. Sheppey},
  \bibinfo{author}{J.~D. Walecka}, \bibinfo{title}{{Neutrino production of
  lepton pairs through the point four-fermion interaction}},
  \bibinfo{journal}{Nuovo Cim.} \bibinfo{volume}{34} (\bibinfo{year}{1964})
  \bibinfo{pages}{404--435}, \bibinfo{doi}{\doi{10.1007/BF02734586}}.

\bibtype{Article}%
\bibitem{Lovseth:1971vv}
\bibinfo{author}{J. Lovseth}, \bibinfo{author}{M. Radomiski},
  \bibinfo{title}{{Kinematical distributions of neutrino-produced lepton
  triplets}}, \bibinfo{journal}{Phys. Rev. D} \bibinfo{volume}{3}
  (\bibinfo{year}{1971}) \bibinfo{pages}{2686--2706},
  \bibinfo{doi}{\doi{10.1103/PhysRevD.3.2686}}.

\bibtype{Article}%
\bibitem{Fujikawa:1971nx}
\bibinfo{author}{K. Fujikawa}, \bibinfo{title}{{The self-coupling of weak
  lepton currents in high-energy neutrino and muon reactions}},
  \bibinfo{journal}{Annals Phys.} \bibinfo{volume}{68} (\bibinfo{year}{1971})
  \bibinfo{pages}{102--162}, \bibinfo{doi}{\doi{10.1016/0003-4916(71)90244-2}}.

\bibtype{Article}%
\bibitem{Koike:1971tu}
\bibinfo{author}{K. Koike}, \bibinfo{author}{M. Konuma}, \bibinfo{author}{K.
  Kurata}, \bibinfo{author}{K. Sugano}, \bibinfo{title}{{Neutrino production of
  lepton pairs. 1. -}}, \bibinfo{journal}{Prog. Theor. Phys.}
  \bibinfo{volume}{46} (\bibinfo{year}{1971}) \bibinfo{pages}{1150--1169},
  \bibinfo{doi}{\doi{10.1143/PTP.46.1150}}.

\bibtype{Article}%
\bibitem{Koike:1971vg}
\bibinfo{author}{K. Koike}, \bibinfo{author}{M. Konuma}, \bibinfo{author}{K.
  Kurata}, \bibinfo{author}{K. Sugano}, \bibinfo{title}{{Neutrino production of
  lepton pairs. 2.}}, \bibinfo{journal}{Prog. Theor. Phys.}
  \bibinfo{volume}{46} (\bibinfo{year}{1971}) \bibinfo{pages}{1799--1804},
  \bibinfo{doi}{\doi{10.1143/PTP.46.1799}}.

\bibtype{Article}%
\bibitem{Brown:1972vne}
\bibinfo{author}{R.~W. Brown}, \bibinfo{author}{R.~H. Hobbs},
  \bibinfo{author}{J. Smith}, \bibinfo{author}{N. Stanko},
  \bibinfo{title}{{Intermediate boson. iii. virtual-boson effects in neutrino
  trident production}}, \bibinfo{journal}{Phys. Rev. D} \bibinfo{volume}{6}
  (\bibinfo{year}{1972}) \bibinfo{pages}{3273--3292},
  \bibinfo{doi}{\doi{10.1103/PhysRevD.6.3273}}.

\bibtype{Article}%
\bibitem{Belusevic:1987cw}
\bibinfo{author}{R. Belusevic}, \bibinfo{author}{J. Smith}, \bibinfo{title}{{W
  - Z Interference in Neutrino - Nucleus Scattering}}, \bibinfo{journal}{Phys.
  Rev. D} \bibinfo{volume}{37} (\bibinfo{year}{1988}) \bibinfo{pages}{2419},
  \bibinfo{doi}{\doi{10.1103/PhysRevD.37.2419}}.

\bibtype{Article}%
\bibitem{Altmannshofer:2014pba}
\bibinfo{author}{Wolfgang Altmannshofer}, \bibinfo{author}{Stefania Gori},
  \bibinfo{author}{Maxim Pospelov}, \bibinfo{author}{Itay Yavin},
  \bibinfo{title}{{Neutrino Trident Production: A Powerful Probe of New Physics
  with Neutrino Beams}}, \bibinfo{journal}{Phys. Rev. Lett.}
  \bibinfo{volume}{113} (\bibinfo{year}{2014}) \bibinfo{pages}{091801},
  \bibinfo{doi}{\doi{10.1103/PhysRevLett.113.091801}}, \eprint{1406.2332}.

\bibtype{Article}%
\bibitem{Magill:2016hgc}
\bibinfo{author}{Gabriel Magill}, \bibinfo{author}{Ryan Plestid},
  \bibinfo{title}{{Neutrino Trident Production at the Intensity Frontier}},
  \bibinfo{journal}{Phys. Rev. D} \bibinfo{volume}{95} (\bibinfo{number}{7})
  (\bibinfo{year}{2017}) \bibinfo{pages}{073004},
  \bibinfo{doi}{\doi{10.1103/PhysRevD.95.073004}}, \eprint{1612.05642}.

\bibtype{Article}%
\bibitem{Ge:2017poy}
\bibinfo{author}{Shao-Feng Ge}, \bibinfo{author}{Manfred Lindner},
  \bibinfo{author}{Werner Rodejohann}, \bibinfo{title}{{Atmospheric Trident
  Production for Probing New Physics}}, \bibinfo{journal}{Phys. Lett. B}
  \bibinfo{volume}{772} (\bibinfo{year}{2017}) \bibinfo{pages}{164--168},
  \bibinfo{doi}{\doi{10.1016/j.physletb.2017.06.020}}, \eprint{1702.02617}.

\bibtype{Article}%
\bibitem{Ballett:2018uuc}
\bibinfo{author}{Peter Ballett}, \bibinfo{author}{Matheus Hostert},
  \bibinfo{author}{Silvia Pascoli}, \bibinfo{author}{Yuber~F. Perez-Gonzalez},
  \bibinfo{author}{Zahra Tabrizi}, \bibinfo{author}{Renata Zukanovich~Funchal},
  \bibinfo{title}{{Neutrino Trident Scattering at Near Detectors}},
  \bibinfo{journal}{JHEP} \bibinfo{volume}{01} (\bibinfo{year}{2019})
  \bibinfo{pages}{119}, \bibinfo{doi}{\doi{10.1007/JHEP01(2019)119}},
  \eprint{1807.10973}.

\bibtype{Article}%
\bibitem{Altmannshofer:2019zhy}
\bibinfo{author}{Wolfgang Altmannshofer}, \bibinfo{author}{Stefania Gori},
  \bibinfo{author}{Justo Mart\'\i{}n-Albo}, \bibinfo{author}{Alexandre Sousa},
  \bibinfo{author}{Michael Wallbank}, \bibinfo{title}{{Neutrino Tridents at
  DUNE}}, \bibinfo{journal}{Phys. Rev. D} \bibinfo{volume}{100}
  (\bibinfo{number}{11}) (\bibinfo{year}{2019}) \bibinfo{pages}{115029},
  \bibinfo{doi}{\doi{10.1103/PhysRevD.100.115029}}, \eprint{1902.06765}.

\bibtype{Article}%
\bibitem{Gauld:2019pgt}
\bibinfo{author}{Rhorry Gauld}, \bibinfo{title}{{Precise predictions for
  multi-TeV and PeV energy neutrino scattering rates}}, \bibinfo{journal}{Phys.
  Rev. D} \bibinfo{volume}{100} (\bibinfo{number}{9}) (\bibinfo{year}{2019})
  \bibinfo{pages}{091301}, \bibinfo{doi}{\doi{10.1103/PhysRevD.100.091301}},
  \eprint{1905.03792}.

\bibtype{Article}%
\bibitem{Zhou:2019vxt}
\bibinfo{author}{Bei Zhou}, \bibinfo{author}{John~F. Beacom},
  \bibinfo{title}{{Neutrino-nucleus cross sections for W-boson and trident
  production}}, \bibinfo{journal}{Phys. Rev. D} \bibinfo{volume}{101}
  (\bibinfo{number}{3}) (\bibinfo{year}{2020}) \bibinfo{pages}{036011},
  \bibinfo{doi}{\doi{10.1103/PhysRevD.101.036011}}, \eprint{1910.08090}.

\bibtype{Article}%
\bibitem{Zhou:2019frk}
\bibinfo{author}{Bei Zhou}, \bibinfo{author}{John~F. Beacom},
  \bibinfo{title}{{W-boson and trident production in TeV\textendash{}PeV
  neutrino observatories}}, \bibinfo{journal}{Phys. Rev. D}
  \bibinfo{volume}{101} (\bibinfo{number}{3}) (\bibinfo{year}{2020})
  \bibinfo{pages}{036010}, \bibinfo{doi}{\doi{10.1103/PhysRevD.101.036010}},
  \eprint{1910.10720}.

\bibtype{Article}%
\bibitem{Altmannshofer:2024hqd}
\bibinfo{author}{Wolfgang Altmannshofer}, \bibinfo{author}{Toni M\"akel\"a},
  \bibinfo{author}{Subir Sarkar}, \bibinfo{author}{Sebastian Trojanowski},
  \bibinfo{author}{Keping Xie}, \bibinfo{author}{Bei Zhou},
  \bibinfo{title}{{Discovering neutrino tridents at the Large Hadron
  Collider}}, \bibinfo{journal}{Phys. Rev. D} \bibinfo{volume}{110}
  (\bibinfo{number}{7}) (\bibinfo{year}{2024}) \bibinfo{pages}{072018},
  \bibinfo{doi}{\doi{10.1103/PhysRevD.110.072018}}, \eprint{2406.16803}.

\bibtype{Article}%
\bibitem{Bigaran:2024zxk}
\bibinfo{author}{Innes Bigaran}, \bibinfo{author}{P.~S.~Bhupal Dev},
  \bibinfo{author}{Diego Lopez~Gutierrez}, \bibinfo{author}{Pedro A.~N.
  Machado}, \bibinfo{title}{{Tau Tridents at Accelerator Neutrino Facilities}}
  (\bibinfo{year}{2024}), \eprint{2406.20067}.

\bibtype{Article}%
\bibitem{Francener:2024wul}
\bibinfo{author}{Reinaldo Francener}, \bibinfo{author}{Victor~P. Goncalves},
  \bibinfo{author}{Diego~R. Gratieri}, \bibinfo{title}{{Neutrino trident
  scattering at the LHC energy regime}}, \bibinfo{journal}{Eur. Phys. J. C}
  \bibinfo{volume}{84} (\bibinfo{number}{9}) (\bibinfo{year}{2024})
  \bibinfo{pages}{923}, \bibinfo{doi}{\doi{10.1140/epjc/s10052-024-13323-2}},
  \eprint{2406.13593}.

\bibtype{Article}%
\bibitem{Seckel:1997kk}
\bibinfo{author}{D. Seckel}, \bibinfo{title}{{Neutrino photon reactions in
  astrophysics and cosmology}}, \bibinfo{journal}{Phys. Rev. Lett.}
  \bibinfo{volume}{80} (\bibinfo{year}{1998}) \bibinfo{pages}{900--903},
  \bibinfo{doi}{\doi{10.1103/PhysRevLett.80.900}}, \eprint{hep-ph/9709290}.

\bibtype{Article}%
\bibitem{Alikhanov:2015kla}
\bibinfo{author}{I. Alikhanov}, \bibinfo{title}{{Hidden Glashow resonance in
  neutrino\textendash{}nucleus collisions}}, \bibinfo{journal}{Phys. Lett. B}
  \bibinfo{volume}{756} (\bibinfo{year}{2016}) \bibinfo{pages}{247--253},
  \bibinfo{doi}{\doi{10.1016/j.physletb.2016.03.009}}, \eprint{1503.08817}.

\bibtype{Article}%
\bibitem{Xie:2023qbn}
\bibinfo{author}{Keping Xie}, \bibinfo{author}{Bei Zhou},
  \bibinfo{author}{T.~J. Hobbs} (\bibinfo{collaboration}{CTEQ-TEA}),
  \bibinfo{title}{{The photon content of the neutron}}, \bibinfo{journal}{JHEP}
  \bibinfo{volume}{04} (\bibinfo{year}{2024}) \bibinfo{pages}{022},
  \bibinfo{doi}{\doi{10.1007/JHEP04(2024)022}}, \eprint{2305.10497}.

\bibtype{Article}%
\bibitem{Ansari:2021cao}
\bibinfo{author}{V. Ansari}, \bibinfo{author}{M.~Sajjad Athar},
  \bibinfo{author}{H. Haider}, \bibinfo{author}{I.~Ruiz Simo},
  \bibinfo{author}{S.~K. Singh}, \bibinfo{author}{F. Zaidi},
  \bibinfo{title}{{Deep inelastic (anti)neutrino\textendash{}nucleus
  scattering}}, \bibinfo{journal}{Eur. Phys. J. ST} \bibinfo{volume}{230}
  (\bibinfo{number}{24}) (\bibinfo{year}{2021}) \bibinfo{pages}{4433--4448},
  \bibinfo{doi}{\doi{10.1140/epjs/s11734-021-00277-9}}, \eprint{2106.14670}.

\bibtype{Article}%
\bibitem{Candido:2023utz}
\bibinfo{author}{Alessandro Candido}, \bibinfo{author}{Alfonso Garcia},
  \bibinfo{author}{Giacomo Magni}, \bibinfo{author}{Tanjona Rabemananjara},
  \bibinfo{author}{Juan Rojo}, \bibinfo{author}{Roy Stegeman},
  \bibinfo{title}{{Neutrino Structure Functions from GeV to EeV Energies}},
  \bibinfo{journal}{JHEP} \bibinfo{volume}{05} (\bibinfo{year}{2023})
  \bibinfo{pages}{149}, \bibinfo{doi}{\doi{10.1007/JHEP05(2023)149}},
  \eprint{2302.08527}.

\bibtype{Article}%
\bibitem{Xie:2023suk}
\bibinfo{author}{Keping Xie}, \bibinfo{author}{Jun Gao}, \bibinfo{author}{T.~J.
  Hobbs}, \bibinfo{author}{Daniel~R. Stump}, \bibinfo{author}{C.~P. Yuan}
  (\bibinfo{collaboration}{CTEQ-TEA}), \bibinfo{title}{{High-energy neutrino
  deep inelastic scattering cross sections}}, \bibinfo{journal}{Phys. Rev. D}
  \bibinfo{volume}{109} (\bibinfo{number}{11}) (\bibinfo{year}{2024})
  \bibinfo{pages}{113001}, \bibinfo{doi}{\doi{10.1103/PhysRevD.109.113001}},
  \eprint{2303.13607}.

\bibtype{Article}%
\bibitem{Jeong:2023hwe}
\bibinfo{author}{Yu~Seon Jeong}, \bibinfo{author}{Mary~Hall Reno},
  \bibinfo{title}{{Neutrino Cross Sections: Interface of shallow- and
  deep-inelastic scattering for collider neutrinos}}  (\bibinfo{year}{2023}),
  \eprint{2307.09241}.

\bibtype{Article}%
\bibitem{Weigel:2024gzh}
\bibinfo{author}{Philip L.~R. Weigel}, \bibinfo{author}{Janet~M. Conrad},
  \bibinfo{author}{Alfonso Garcia-Soto}, \bibinfo{title}{{Cross sections and
  inelasticity distributions of high-energy neutrino deep inelastic
  scattering}}, \bibinfo{journal}{Phys. Rev. D} \bibinfo{volume}{111}
  (\bibinfo{number}{4}) (\bibinfo{year}{2025}) \bibinfo{pages}{043044},
  \bibinfo{doi}{\doi{10.1103/PhysRevD.111.043044}}, \eprint{2408.05866}.

\bibtype{Article}%
\bibitem{FerrarioRavasio:2024kem}
\bibinfo{author}{Silvia Ferrario~Ravasio}, \bibinfo{author}{Rhorry Gauld},
  \bibinfo{author}{Barbara J\"ager}, \bibinfo{author}{Alexander Karlberg},
  \bibinfo{author}{Giulia Zanderighi}, \bibinfo{title}{{An event generator for
  neutrino-induced Deep Inelastic Scattering and applications to neutrino
  astronomy}}  (\bibinfo{year}{2024}), \eprint{2407.03894}.

\bibtype{Article}%
\bibitem{vanBeekveld:2024ziz}
\bibinfo{author}{Melissa van Beekveld}, \bibinfo{author}{Silvia
  Ferrario~Ravasio}, \bibinfo{author}{Eva Groenendijk}, \bibinfo{author}{Peter
  Krack}, \bibinfo{author}{Juan Rojo}, \bibinfo{author}{Valentina~Sch\"utze
  S\'anchez}, \bibinfo{title}{{A phenomenological analysis of LHC neutrino
  scattering at NLO accuracy matched to parton showers}},
  \bibinfo{journal}{Eur. Phys. J. C} \bibinfo{volume}{84}
  (\bibinfo{number}{11}) (\bibinfo{year}{2024}) \bibinfo{pages}{1175},
  \bibinfo{doi}{\doi{10.1140/epjc/s10052-024-13386-1}}, \eprint{2407.09611}.

\bibtype{Article}%
\bibitem{COHERENT:2017ipa}
\bibinfo{author}{D. Akimov}, et al. (\bibinfo{collaboration}{COHERENT}),
  \bibinfo{title}{{Observation of Coherent Elastic Neutrino-Nucleus
  Scattering}}, \bibinfo{journal}{Science} \bibinfo{volume}{357}
  (\bibinfo{number}{6356}) (\bibinfo{year}{2017}) \bibinfo{pages}{1123--1126},
  \bibinfo{doi}{\doi{10.1126/science.aao0990}}, \eprint{1708.01294}.

\bibtype{Article}%
\bibitem{Brdar:2021hpy}
\bibinfo{author}{Vedran Brdar}, \bibinfo{author}{Andr\'e de Gouv\^ea},
  \bibinfo{author}{Pedro A.~N. Machado}, \bibinfo{author}{Ryan Plestid},
  \bibinfo{title}{{Resonances in \ensuremath{\nu}\textasciimacron{}e-e-
  scattering below a TeV}}, \bibinfo{journal}{Phys. Rev. D}
  \bibinfo{volume}{105} (\bibinfo{number}{9}) (\bibinfo{year}{2022})
  \bibinfo{pages}{093004}, \bibinfo{doi}{\doi{10.1103/PhysRevD.105.093004}},
  \eprint{2112.03283}.

\bibtype{Article}%
\bibitem{Plestid:2024bva}
\bibinfo{author}{Ryan Plestid}, \bibinfo{author}{Bei Zhou},
  \bibinfo{title}{{Final state radiation from high and ultrahigh energy
  neutrino interactions}}  (\bibinfo{year}{2024}), \eprint{2403.07984}.

\bibtype{Article}%
\bibitem{Ackermann:2022rqc}
\bibinfo{author}{Markus Ackermann}, et al., \bibinfo{title}{{High-energy and
  ultra-high-energy neutrinos: A Snowmass white paper}},
  \bibinfo{journal}{JHEAp} \bibinfo{volume}{36} (\bibinfo{year}{2022})
  \bibinfo{pages}{55--110}, \bibinfo{doi}{\doi{10.1016/j.jheap.2022.08.001}},
  \eprint{2203.08096}.

\bibtype{Article}%
\bibitem{DeLellis:2004ovi}
\bibinfo{author}{Giovanni De~Lellis}, \bibinfo{author}{Pasquale Migliozzi},
  \bibinfo{author}{Pietro Santorelli}, \bibinfo{title}{{Charm physics with
  neutrinos}}, \bibinfo{journal}{Phys. Rept.} \bibinfo{volume}{399}
  (\bibinfo{year}{2004}) \bibinfo{pages}{227--320},
  \bibinfo{doi}{\doi{10.1016/j.physrep.2005.02.001}}, \bibinfo{note}{[Erratum:
  Phys.Rept. 411, 323--324 (2005)]}.

\bibtype{Article}%
\bibitem{Hou:2019efy}
\bibinfo{author}{Tie-Jiun Hou}, et al., \bibinfo{title}{{New CTEQ global
  analysis of quantum chromodynamics with high-precision data from the LHC}},
  \bibinfo{journal}{Phys. Rev. D} \bibinfo{volume}{103} (\bibinfo{number}{1})
  (\bibinfo{year}{2021}) \bibinfo{pages}{014013},
  \bibinfo{doi}{\doi{10.1103/PhysRevD.103.014013}}, \eprint{1912.10053}.

\bibtype{Article}%
\bibitem{Faura:2020oom}
\bibinfo{author}{Ferran Faura}, \bibinfo{author}{Shayan Iranipour},
  \bibinfo{author}{Emanuele~R. Nocera}, \bibinfo{author}{Juan Rojo},
  \bibinfo{author}{Maria Ubiali}, \bibinfo{title}{{The Strangest Proton?}},
  \bibinfo{journal}{Eur. Phys. J. C} \bibinfo{volume}{80}
  (\bibinfo{number}{12}) (\bibinfo{year}{2020}) \bibinfo{pages}{1168},
  \bibinfo{doi}{\doi{10.1140/epjc/s10052-020-08749-3}}, \eprint{2009.00014}.

\bibtype{Article}%
\bibitem{Zhou:2021xuh}
\bibinfo{author}{Bei Zhou}, \bibinfo{author}{John~F. Beacom},
  \bibinfo{title}{{Dimuons in neutrino telescopes: New predictions and first
  search in IceCube}}, \bibinfo{journal}{Phys. Rev. D} \bibinfo{volume}{105}
  (\bibinfo{number}{9}) (\bibinfo{year}{2022}) \bibinfo{pages}{093005},
  \bibinfo{doi}{\doi{10.1103/PhysRevD.105.093005}}, \eprint{2110.02974}.

\bibtype{Article}%
\bibitem{FASER:2018eoc}
\bibinfo{author}{Akitaka Ariga}, et al. (\bibinfo{collaboration}{FASER}),
  \bibinfo{title}{{FASER\textquoteright{}s physics reach for long-lived
  particles}}, \bibinfo{journal}{Phys. Rev. D} \bibinfo{volume}{99}
  (\bibinfo{number}{9}) (\bibinfo{year}{2019}) \bibinfo{pages}{095011},
  \bibinfo{doi}{\doi{10.1103/PhysRevD.99.095011}}, \eprint{1811.12522}.

\bibtype{Article}%
\bibitem{Cheung:2023gwm}
\bibinfo{author}{Kingman Cheung}, \bibinfo{author}{Thong T.~Q. Nguyen},
  \bibinfo{author}{C.~J. Ouseph}, \bibinfo{title}{{Leptoquark search at the
  Forward Physics Facility}}, \bibinfo{journal}{Phys. Rev. D}
  \bibinfo{volume}{108} (\bibinfo{number}{3}) (\bibinfo{year}{2023})
  \bibinfo{pages}{036014}, \bibinfo{doi}{\doi{10.1103/PhysRevD.108.036014}},
  \eprint{2302.05461}.

\bibtype{Article}%
\bibitem{FASER:2023tle}
\bibinfo{author}{Henso Abreu}, et al. (\bibinfo{collaboration}{FASER}),
  \bibinfo{title}{{Search for dark photons with the FASER detector at the
  LHC}}, \bibinfo{journal}{Phys. Lett. B} \bibinfo{volume}{848}
  (\bibinfo{year}{2024}) \bibinfo{pages}{138378},
  \bibinfo{doi}{\doi{10.1016/j.physletb.2023.138378}}, \eprint{2308.05587}.

\bibtype{Article}%
\bibitem{FASER:2024bbl}
\bibinfo{author}{Roshan Mammen~Abraham}, et al.
  (\bibinfo{collaboration}{FASER}), \bibinfo{title}{{Shining light on the dark
  sector: search for axion-like particles and other new physics in photonic
  final states with FASER}}, \bibinfo{journal}{JHEP} \bibinfo{volume}{01}
  (\bibinfo{year}{2025}) \bibinfo{pages}{199},
  \bibinfo{doi}{\doi{10.1007/JHEP01(2025)199}}, \eprint{2410.10363}.

\bibtype{Article}%
\bibitem{MammenAbraham:2024gun}
\bibinfo{author}{Roshan Mammen~Abraham}, \bibinfo{author}{Jyotismita Adhikary},
  \bibinfo{author}{Jonathan~L. Feng}, \bibinfo{author}{Max Fieg},
  \bibinfo{author}{Felix Kling}, \bibinfo{author}{Jinmian Li},
  \bibinfo{author}{Junle Pei}, \bibinfo{author}{Tanjona~R. Rabemananjara},
  \bibinfo{author}{Juan Rojo}, \bibinfo{author}{Sebastian Trojanowski},
  \bibinfo{title}{{FPF@FCC: Neutrino, QCD, and BSM Physics Opportunities with
  Far-Forward Experiments at a 100 TeV Proton Collider}}
  (\bibinfo{year}{2024}), \eprint{2409.02163}.

\bibtype{Article}%
\bibitem{Ismail:2020yqc}
\bibinfo{author}{Ahmed Ismail}, \bibinfo{author}{Roshan Mammen~Abraham},
  \bibinfo{author}{Felix Kling}, \bibinfo{title}{{Neutral current neutrino
  interactions at FASER$\nu$}}, \bibinfo{journal}{Phys. Rev. D}
  \bibinfo{volume}{103} (\bibinfo{number}{5}) (\bibinfo{year}{2021})
  \bibinfo{pages}{056014}, \bibinfo{doi}{\doi{10.1103/PhysRevD.103.056014}},
  \eprint{2012.10500}.

\bibtype{Article}%
\bibitem{Falkowski:2021bkq}
\bibinfo{author}{Adam Falkowski}, \bibinfo{author}{Mart\'\i{}n
  Gonz\'alez-Alonso}, \bibinfo{author}{Joachim Kopp}, \bibinfo{author}{Yotam
  Soreq}, \bibinfo{author}{Zahra Tabrizi}, \bibinfo{title}{{EFT at
  FASER\ensuremath{\nu}}}, \bibinfo{journal}{JHEP} \bibinfo{volume}{10}
  (\bibinfo{year}{2021}) \bibinfo{pages}{086},
  \bibinfo{doi}{\doi{10.1007/JHEP10(2021)086}}, \eprint{2105.12136}.

\bibtype{Article}%
\bibitem{Kling:2023tgr}
\bibinfo{author}{Felix Kling}, \bibinfo{author}{Toni M\"akel\"a},
  \bibinfo{author}{Sebastian Trojanowski}, \bibinfo{title}{{Investigating the
  fluxes and physics potential of LHC neutrino experiments}},
  \bibinfo{journal}{Phys. Rev. D} \bibinfo{volume}{108} (\bibinfo{number}{9})
  (\bibinfo{year}{2023}) \bibinfo{pages}{095020},
  \bibinfo{doi}{\doi{10.1103/PhysRevD.108.095020}}, \eprint{2309.10417}.

\bibtype{Article}%
\bibitem{IceCube:2013low}
\bibinfo{author}{M.~G. Aartsen}, et al. (\bibinfo{collaboration}{IceCube}),
  \bibinfo{title}{{Evidence for High-Energy Extraterrestrial Neutrinos at the
  IceCube Detector}}, \bibinfo{journal}{Science} \bibinfo{volume}{342}
  (\bibinfo{year}{2013}) \bibinfo{pages}{1242856},
  \bibinfo{doi}{\doi{10.1126/science.1242856}}, \eprint{1311.5238}.

\bibtype{Article}%
\bibitem{Senno:2015tsn}
\bibinfo{author}{Nicholas Senno}, \bibinfo{author}{Kohta Murase},
  \bibinfo{author}{Peter Meszaros}, \bibinfo{title}{{Choked Jets and
  Low-Luminosity Gamma-Ray Bursts as Hidden Neutrino Sources}},
  \bibinfo{journal}{Phys. Rev. D} \bibinfo{volume}{93} (\bibinfo{number}{8})
  (\bibinfo{year}{2016}) \bibinfo{pages}{083003},
  \bibinfo{doi}{\doi{10.1103/PhysRevD.93.083003}}, \eprint{1512.08513}.

\bibtype{Article}%
\bibitem{Senno:2017vtd}
\bibinfo{author}{Nicholas Senno}, \bibinfo{author}{Kohta Murase},
  \bibinfo{author}{Peter Mészáros}, \bibinfo{title}{{Constraining high-energy
  neutrino emission from choked jets in stripped-envelope supernovae}},
  \bibinfo{journal}{JCAP} \bibinfo{volume}{1801} (\bibinfo{year}{2018})
  \bibinfo{pages}{025}, \bibinfo{doi}{\doi{10.1088/1475-7516/2018/01/025}},
  \eprint{1706.02175}.

\bibtype{Article}%
\bibitem{Esmaili:2018wnv}
\bibinfo{author}{Arman Esmaili}, \bibinfo{author}{Kohta Murase},
  \bibinfo{title}{{Constraining high-energy neutrinos from choked-jet
  supernovae with IceCube high-energy starting events}},
  \bibinfo{journal}{JCAP} \bibinfo{volume}{1812} (\bibinfo{year}{2018})
  \bibinfo{pages}{008}, \bibinfo{doi}{\doi{10.1088/1475-7516/2018/12/008}},
  \eprint{1809.09610}.

\bibtype{Article}%
\bibitem{Chang:2022hqj}
\bibinfo{author}{Po-Wen Chang}, \bibinfo{author}{Bei Zhou},
  \bibinfo{author}{Kohta Murase}, \bibinfo{author}{Marc Kamionkowski},
  \bibinfo{title}{{High-energy neutrinos from choked-jet supernovae: Searches
  and implications}}, \bibinfo{journal}{Phys. Rev. D} \bibinfo{volume}{109}
  (\bibinfo{number}{10}) (\bibinfo{year}{2024}) \bibinfo{pages}{103041},
  \bibinfo{doi}{\doi{10.1103/PhysRevD.109.103041}}, \eprint{2210.03088}.

\bibtype{Article}%
\bibitem{IceCube:2023esf}
\bibinfo{author}{R. Abbasi}, et al. (\bibinfo{collaboration}{IceCube}),
  \bibinfo{title}{{Constraining High-energy Neutrino Emission from Supernovae
  with IceCube}}, \bibinfo{journal}{Astrophys. J. Lett.} \bibinfo{volume}{949}
  (\bibinfo{number}{1}) (\bibinfo{year}{2023}) \bibinfo{pages}{L12},
  \bibinfo{doi}{\doi{10.3847/2041-8213/acd2c9}}, \eprint{2303.03316}.

\bibtype{Article}%
\bibitem{Gaisser:2002jj}
\bibinfo{author}{T.~K. Gaisser}, \bibinfo{author}{M. Honda},
  \bibinfo{title}{{Flux of atmospheric neutrinos}}, \bibinfo{journal}{Ann. Rev.
  Nucl. Part. Sci.} \bibinfo{volume}{52} (\bibinfo{year}{2002})
  \bibinfo{pages}{153--199},
  \bibinfo{doi}{\doi{10.1146/annurev.nucl.52.050102.090645}},
  \eprint{hep-ph/0203272}.

\bibtype{Misc}%
\bibitem{IceCube_web}
\bibinfo{howpublished}{\url{https://icecube.wisc.edu/}}.

\bibtype{Article}%
\bibitem{KM3Net:2016zxf}
\bibinfo{author}{S. Adrian-Martinez}, et al. (\bibinfo{collaboration}{KM3Net}),
  \bibinfo{title}{{Letter of intent for KM3NeT 2.0}}, \bibinfo{journal}{J.
  Phys. G} \bibinfo{volume}{43} (\bibinfo{number}{8}) (\bibinfo{year}{2016})
  \bibinfo{pages}{084001}, \bibinfo{doi}{\doi{10.1088/0954-3899/43/8/084001}},
  \eprint{1601.07459}.

\bibtype{Article}%
\bibitem{Allakhverdyan:2021vkk}
\bibinfo{author}{V.~A. Allakhverdyan}, et al., \bibinfo{title}{{Deep-Water
  Neutrino Telescope in Lake Baikal}}, \bibinfo{journal}{Phys. At. Nucl.}
  \bibinfo{volume}{84} (\bibinfo{number}{9}) (\bibinfo{year}{2021})
  \bibinfo{pages}{1600--1609}, \bibinfo{doi}{\doi{10.1134/S1063778821090064}}.

\bibtype{Article}%
\bibitem{IceCube-Gen2:2020qha}
\bibinfo{author}{M.~G. Aartsen}, et al.
  (\bibinfo{collaboration}{IceCube-Gen2}), \bibinfo{title}{{IceCube-Gen2: the
  window to the extreme Universe}}, \bibinfo{journal}{J. Phys. G}
  \bibinfo{volume}{48} (\bibinfo{number}{6}) (\bibinfo{year}{2021})
  \bibinfo{pages}{060501}, \bibinfo{doi}{\doi{10.1088/1361-6471/abbd48}},
  \eprint{2008.04323}.

\bibtype{Misc}%
\bibitem{Gen2_TDR_web}
\bibinfo{howpublished}{\url{https://icecube-gen2.wisc.edu/science/publications/TDR/}}.

\bibtype{Article}%
\bibitem{P-ONE:2020ljt}
\bibinfo{author}{Matteo Agostini}, et al. (\bibinfo{collaboration}{P-ONE}),
  \bibinfo{title}{{The Pacific Ocean Neutrino Experiment}},
  \bibinfo{journal}{Nature Astron.} \bibinfo{volume}{4} (\bibinfo{number}{10})
  (\bibinfo{year}{2020}) \bibinfo{pages}{913--915},
  \bibinfo{doi}{\doi{10.1038/s41550-020-1182-4}}, \eprint{2005.09493}.

\bibtype{Article}%
\bibitem{Ye:2023dch}
\bibinfo{author}{Z.~P. Ye}, et al., \bibinfo{title}{{A multi-cubic-kilometre
  neutrino telescope in the western Pacific Ocean}}, \bibinfo{journal}{Nature
  Astron.} \bibinfo{volume}{7} (\bibinfo{number}{12}) (\bibinfo{year}{2023})
  \bibinfo{pages}{1497--1505}, \bibinfo{doi}{\doi{10.1038/s41550-023-02087-6}}.

\bibtype{Article}%
\bibitem{Huang:2023mzt}
\bibinfo{author}{Tian-Qi Huang}, \bibinfo{author}{Zhen Cao},
  \bibinfo{author}{Mingjun Chen}, \bibinfo{author}{Jiali Liu},
  \bibinfo{author}{Zike Wang}, \bibinfo{author}{Xiaohao You},
  \bibinfo{author}{Ying Qi}, \bibinfo{title}{{Proposal for the High Energy
  Neutrino Telescope}}, \bibinfo{journal}{PoS} \bibinfo{volume}{ICRC2023}
  (\bibinfo{year}{2023}) \bibinfo{pages}{1080},
  \bibinfo{doi}{\doi{10.22323/1.444.1080}}.

\bibtype{Article}%
\bibitem{Zhang:2024slv}
\bibinfo{author}{Huiming Zhang}, \bibinfo{author}{Yudong Cui},
  \bibinfo{author}{Yunlei Huang}, \bibinfo{author}{Sujie Lin},
  \bibinfo{author}{Yihan Liu}, \bibinfo{author}{Zijian Qiu},
  \bibinfo{author}{Chengyu Shao}, \bibinfo{author}{Yihan Shi},
  \bibinfo{author}{Caijin Xie}, \bibinfo{author}{Lili Yang}, \bibinfo{title}{{A
  proposed deep sea Neutrino Observatory in the Nanhai}},
  \bibinfo{journal}{Astropart. Phys.} \bibinfo{volume}{171}
  (\bibinfo{year}{2025}) \bibinfo{pages}{103123},
  \bibinfo{doi}{\doi{10.1016/j.astropartphys.2025.103123}},
  \eprint{2408.05122}.

\bibtype{Article}%
\bibitem{Fieg:2023kld}
\bibinfo{author}{Max Fieg}, \bibinfo{author}{Felix Kling},
  \bibinfo{author}{Holger Schulz}, \bibinfo{author}{Torbj\"orn Sj\"ostrand},
  \bibinfo{title}{{Tuning Pythia for Forward Physics Experiments}}
  (\bibinfo{year}{2023}), \eprint{2309.08604}.

\bibtype{Article}%
\bibitem{Soldin:2021wyv}
\bibinfo{author}{Dennis Soldin} (\bibinfo{collaboration}{EAS-MSU, IceCube,
  KASCADE-Grande, NEVOD-DECOR, Pierre Auger, SUGAR, Telescope Array, Yakutsk
  EAS Array}), \bibinfo{title}{{Update on the Combined Analysis of Muon
  Measurements from Nine Air Shower Experiments}}, \bibinfo{journal}{PoS}
  \bibinfo{volume}{ICRC2021} (\bibinfo{year}{2021}) \bibinfo{pages}{349},
  \bibinfo{doi}{\doi{10.22323/1.395.0349}}, \eprint{2108.08341}.

\bibtype{Article}%
\bibitem{Farrar:2013sfa}
\bibinfo{author}{Glennys~R. Farrar}, \bibinfo{author}{Jeffrey~D. Allen},
  \bibinfo{title}{{A new physical phenomenon in ultra-high energy collisions}},
  \bibinfo{journal}{EPJ Web Conf.} \bibinfo{volume}{53} (\bibinfo{year}{2013})
  \bibinfo{pages}{07007}, \bibinfo{doi}{\doi{10.1051/epjconf/20135307007}},
  \eprint{1307.2322}.

\bibtype{Article}%
\bibitem{Anchordoqui:2016oxy}
\bibinfo{author}{Luis~A. Anchordoqui}, \bibinfo{author}{Haim Goldberg},
  \bibinfo{author}{Thomas~J. Weiler}, \bibinfo{title}{{Strange fireball as an
  explanation of the muon excess in Auger data}}, \bibinfo{journal}{Phys. Rev.
  D} \bibinfo{volume}{95} (\bibinfo{number}{6}) (\bibinfo{year}{2017})
  \bibinfo{pages}{063005}, \bibinfo{doi}{\doi{10.1103/PhysRevD.95.063005}},
  \eprint{1612.07328}.

\bibtype{Article}%
\bibitem{Baur:2019cpv}
\bibinfo{author}{Sebastian Baur}, \bibinfo{author}{Hans Dembinski},
  \bibinfo{author}{Matias Perlin}, \bibinfo{author}{Tanguy Pierog},
  \bibinfo{author}{Ralf Ulrich}, \bibinfo{author}{Klaus Werner},
  \bibinfo{title}{{Core-corona effect in hadron collisions and muon production
  in air showers}}, \bibinfo{journal}{Phys. Rev. D} \bibinfo{volume}{107}
  (\bibinfo{number}{9}) (\bibinfo{year}{2023}) \bibinfo{pages}{094031},
  \bibinfo{doi}{\doi{10.1103/PhysRevD.107.094031}}, \eprint{1902.09265}.

\bibtype{Article}%
\bibitem{Anchordoqui:2019laz}
\bibinfo{author}{Luis~A. Anchordoqui}, \bibinfo{author}{Carlos
  Garc\'\i{}a~Canal}, \bibinfo{author}{Sergio~J. Sciutto},
  \bibinfo{author}{Jorge~F. Soriano}, \bibinfo{title}{{Through the
  looking-glass with ALICE into the quark-gluon plasma: A new test for hadronic
  interaction models used in air shower simulations}}, \bibinfo{journal}{Phys.
  Lett. B} \bibinfo{volume}{810} (\bibinfo{year}{2020})
  \bibinfo{pages}{135837}, \bibinfo{doi}{\doi{10.1016/j.physletb.2020.135837}},
  \eprint{1907.09816}.

\bibtype{Article}%
\bibitem{Pierog:2020ghc}
\bibinfo{author}{Tanguy Pierog}, \bibinfo{author}{Sebastian Baur},
  \bibinfo{author}{Hans~P. Dembinski}, \bibinfo{author}{Ralf Ulrich},
  \bibinfo{author}{Klaus Werner}, \bibinfo{title}{{Collective Hadronization and
  Air Showers: Can LHC Data Solve the Muon Puzzle ?}}, \bibinfo{journal}{PoS}
  \bibinfo{volume}{ICRC2019} (\bibinfo{year}{2020}) \bibinfo{pages}{387},
  \bibinfo{doi}{\doi{10.22323/1.358.0387}}.

\bibtype{Article}%
\bibitem{Ostapchenko:2013pia}
\bibinfo{author}{S. Ostapchenko}, \bibinfo{title}{{QGSJET-II: physics, recent
  improvements, and results for air showers}}, \bibinfo{journal}{EPJ Web Conf.}
  \bibinfo{volume}{52} (\bibinfo{year}{2013}) \bibinfo{pages}{02001},
  \bibinfo{doi}{\doi{10.1051/epjconf/20125202001}}.

\bibtype{Article}%
\bibitem{Riehn:2019jet}
\bibinfo{author}{Felix Riehn}, \bibinfo{author}{Ralph Engel},
  \bibinfo{author}{Anatoli Fedynitch}, \bibinfo{author}{Thomas~K. Gaisser},
  \bibinfo{author}{Todor Stanev}, \bibinfo{title}{{Hadronic interaction model
  Sibyll 2.3d and extensive air showers}}, \bibinfo{journal}{Phys. Rev. D}
  \bibinfo{volume}{102} (\bibinfo{number}{6}) (\bibinfo{year}{2020})
  \bibinfo{pages}{063002}, \bibinfo{doi}{\doi{10.1103/PhysRevD.102.063002}},
  \eprint{1912.03300}.

\bibtype{Article}%
\bibitem{Fedynitch:2115393}
\bibinfo{author}{A Fedynitch}, \bibinfo{author}{R Engel},
  \bibinfo{title}{{Revision of the high energy hadronic interaction models
  PHOJET/DPMJET-III}}  (\bibinfo{year}{2015}),
  \bibinfo{url}{\urlprefix\url{https://cds.cern.ch/record/2115393}}.

\bibtype{Article}%
\bibitem{Anchordoqui:2022fpn}
\bibinfo{author}{Luis~A. Anchordoqui}, \bibinfo{author}{Carlos~Garcia Canal},
  \bibinfo{author}{Felix Kling}, \bibinfo{author}{Sergio~J. Sciutto},
  \bibinfo{author}{Jorge~F. Soriano}, \bibinfo{title}{{An explanation of the
  muon puzzle of ultrahigh-energy cosmic rays and the role of the Forward
  Physics Facility for model improvement}}, \bibinfo{journal}{JHEAp}
  \bibinfo{volume}{34} (\bibinfo{year}{2022}) \bibinfo{pages}{19--32},
  \bibinfo{doi}{\doi{10.1016/j.jheap.2022.03.004}}, \eprint{2202.03095}.

\end{thebibliography*}

\end{document}